\documentclass{ieeeaccess}
\usepackage{cite}
\usepackage{amsmath,amssymb,amsfonts}
\usepackage{algorithmic}
\usepackage{graphicx}
\usepackage{textcomp}
\usepackage{multirow}
\usepackage{graphicx}
\usepackage{placeins}
\usepackage[hyphens]{url}
\usepackage[caption=false]{subfig}
\usepackage{caption,setspace}
\def\BibTeX{{\rm B\kern-.05em{\sc i\kern-.025em b}\kern-.08em
    T\kern-.1667em\lower.7ex\hbox{E}\kern-.125emX}}
    
\usepackage[table]{xcolor}
\usepackage{tabularx}
\usepackage{amssymb}% http://ctan.org/pkg/amssymb
\usepackage{pifont}% http://ctan.org/pkg/pifont
\newcommand{\cmark}{\ding{51}}%
\newcommand{\xmark}{\ding{55}}%

\newcolumntype{s}{>{\columncolor[HTML]{AAACED}} p{3cm}}
\arrayrulecolor[HTML]{211d1b}

\begin{document}
\history{Date of publication xxxx 00, 0000, date of current version xxxx 00, 0000.}
\doi{10.1109/ACCESS.2017.DOI}

% \title{Healthcare with Distributed Ledger Technology Solution}
\title{Distributed Ledger Technology based Integrated Healthcare Solution for Bangladesh}
\author{\uppercase{Md. Ariful Islam}\authorrefmark{1} \IEEEmembership{Member, IEEE},
\uppercase{Md. Antonin Islam\authorrefmark{2}, \uppercase{Md. Amzad Hossain Jacky}\authorrefmark{2}, Md. Al-Amin\authorrefmark{2}, M. Saef Ullah Miah\authorrefmark{3} \IEEEmembership{Member, IEEE}, Md Muhidul Islam Khan\authorrefmark{4} and Md. Iqbal Hossain\authorrefmark{5}}}

\address[1]{Brain Station 23 Ltd., Dhaka, Bangladesh}
\address[2]{American International University-Bangladesh, Dhaka, Bangladesh}
\address[3]{Faculty of Computing, College of Computing and Applied Sciences, Universiti Malaysia Pahang, Pekan 26600, Malaysia}
\address[4]{Department of Electronics and Computer Science, University of Stavanger, Stavanger, Norway}
\address[5]{Department of Paediatrics, Chakaria Unique Hospital,  Chattogram, Bangladesh}

% \tfootnote{This paragraph of the first footnote will contain support 
% information, including sponsor and financial support acknowledgment. For 
% example, ``This work was supported in part by the U.S. Department of 
% Commerce under Grant BS123456.''}

\markboth
{Ariful et.al, \headeretal: Distributed Ledger Technology based Integrated Healthcare Solution for Bangladesh}
{Ariful et.al, \headeretal: Distributed Ledger Technology based Integrated Healthcare Solution for Bangladesh}

\corresp{Corresponding author: Md. Ariful Islam (e-mail: fahim.arif0373@outlook.com).}

\begin{abstract}

Healthcare data is sensitive and requires great protection. Encrypted electronic health records (EHRs) contain personal and sensitive data such as names and addresses. Having access to patient data benefits all of them. This paper proposes a blockchain-based distributed healthcare application platform for Bangladeshi public and private healthcare providers. Using data immutability and smart contracts, the suggested application framework allows users to create safe digital agreements for commerce or collaboration. Thus, all enterprises may securely collaborate using the same blockchain network, gaining data openness and read/write capacity. The proposed application consists of various application interfaces for various system users. For data integrity, privacy, permission and service availability, the proposed solution leverages Hyperledger fabric and Blockchain as a Service. Everyone will also have their own profile in the portal. A unique identity for each person and the installation of digital information centres across the country have greatly eased the process. It will collect systematic health data from each person which will be beneficial for research institutes and health-related organisations. A national data warehouse in Bangladesh is feasible for this application and It is also possible to keep a clean health sector by analysing data stored in this warehouse and conducting various purification algorithms using technologies like Data Science. Given that Bangladesh has both public and private health care, a straightforward digital strategy for all organisations is essential.

\end{abstract}


\titlepgskip=-15pt

\maketitle

\section{Introduction}
\label{sec:introduction}
\PARstart{T}{he} healthcare industry is an important part of a country's economy that includes many medical products and services that are also basic necessities of life. This sector includes an extensive chain of services to which the government contributes in a big way. For example, the government of the People's Republic of Bangladesh has at least 104,659 human resources for health services~\cite{1} for which the government budgeted \$2,174 million~\cite{1} in 2007, and the budget is increasing day by day according to the public demand. The People's Republic of Bangladesh offers a comprehensive Essential Health Service Package (ESP), which includes many free medicines of various generic drugs and also offers many low cost medical tests which are presented in Table \ref{table:minimumStandards}. However, as these services are provided in an area of 148,460 sq km, densely populated with 163 million people (2019, World Bank), many accuse the country of mismanagement, imbalance and lack of service distribution. As a result, the national public health service cannot properly meet public satisfaction~\cite{3}.

To address all these shortcomings, the government is trying to find technology-based solutions, and many third-party companies are trying to solve these problems on behalf of the government. However, since it is a nationwide service with a large number of administrators, the government should move the entire administrative system to a digital platform where every action (only the official one) of each health official is counted as data, and all data can be authenticated, stored and analyzed for monitoring and management. In this way, Bangladesh can take another step towards digitization.

But with this great philosophy of digitizing a nation comes with a responsibility for digital security (data, identity, assets) and cybersecurity (information and IT infrastructure). After all, when an organization or institution deals with physical documents, many security measures can be taken. The global monetary system is a good example of this. These can have so many factual printing patterns that they can be easily distinguished from other counterfeit copies. But when it comes to digital data, tampering can be done from the central database without a trace. This can lead to a national or international threat. The Kaseya Ransomware attack~\cite{oxfordkaseya} is a good example of this. It affected over 1500 companies and over 1 million locked computers. So as the world moves towards a world war III that will not be a war of arms but primarily a cyberwar, it is a major concern for decision-makers in ministries and organizations to approve most official functions online.

At the same time, the ongoing trial is fraught with so many allegations. Some are aimed directly at officials for lack of proper oversight and system gaps, and most are due to an attempt to serve a populous nation with traditional monolithic management.

Adopting a digital platform in the ministerial office environment is not a new challenge for Bangladesh. Central and state-owned banks are good example. They work with global banking systems and at the same time with local managed file transfer (MFT) and payment gateway services. However, the issue is that of security, traceability and immutability of digital data. Therefore, the importance of the new cutting-edge technology known as Blockchain cannot be overlooked. In order to introduce this emerging technology at a national level, we need to promote it at an early stage. Therefore, the solution proposed in this article can be a breakthrough in the movement of the world's fourth industrial revolution~\cite{miah2020geofencing}. The proposed solution is mainly about proposing a digital information management system that ensures the status of the public health service, transparency of information and security.

More specifically, the project is a complete solution designed to provide a single service to the general population of Bangladesh. The back-end will be protected with blockchain-based decentralized ledger technology. So, there will be no dependence on third party for digital healthcare. Due to technical limitations, no public or private medical complex or hospital will be left behind in service delivery . All healthcare providers will be connected as peers to a single digital health service at the national level. So competition between different medical facilities for good services will continue, but the population will access these services from a single source through their citizen health portal. In addition to solving all these current problems of civilian service, our project can be an important tool in the near future to improve civilian life in more diverse ways.

A decentralized blockchain service might be the best choice with several governmental and non-governmental organizations involved in clinical / biomedical research. Using this technology, different organizations can collaborate with other institutions/organizations to share and analyze data without relinquishing control. Each institution can maintain full control over its own computing resources while collaborating with other institutions. In this way, a robust dataset can be formed for each clinical trial \cite{ichikawa2017tamper}. Fig. \ref{fig:ehrCycle} provides an overview of this idea of secure collaboration. Patient-generated data becomes available to researchers. As a result, a decent environment ,as shown in Fig. \ref{fig:ehrCycle} can be created where academia, healthcare industry and healthcare professionals can collaborate. In case of security and privacy, private blockchain can be an excellent option to control transparency, security ,and immutability of data. If anonymized and then tracked in the research process with a timestamp, this secondary data source would enable millions of individuals, healthcare providers, healthcare institutions, and medical researchers to share vast amounts of genetic, nutritional, lifestyle, environmental, and health data with guaranteed security and privacy \cite{ichikawa2017tamper}.

\begin{table*}[!htbp]

\centering
	\resizebox{\textwidth}{!}{\begin{tabular}{|c| c| c| c| c|} 
%\rowcolor{lightgray} 
\hline
\textbf{Community Clinic} & \textbf{Union Health and Family Welfare Centre} & \textbf{Upazila Health Complex} & \textbf{Maternal and Child Welfare Centre} & \textbf{District Hospital} \\
\hline

 & & & & Trauma Care\\

 & & & & Ophthalm. Surgery\\
 
 & & General Surgery & & General Surgery\\
 
 & & Obstetric Fistula & & Obstetric Fistula\\

 & & CEmONC & & \cellcolor[HTML]{F7C9AB}CEmONC \\

 & & \cellcolor[HTML]{F7C9AB}Severe cases & & \cellcolor[HTML]{F7C9AB}Severe cases\\
 
 & BEmONC & \cellcolor[HTML]{F7C9AB}BEmONC & CEmONC & \cellcolor[HTML]{F7C9AB}BEmONC\\

Normal Newborn & \cellcolor[HTML]{F7C9AB}Normal Newborn & \cellcolor[HTML]{F7C9AB}Normal Newborn & \cellcolor[HTML]{F7C9AB}Pre-term NB & \cellcolor[HTML]{F7C9AB}Normal Newborn \\

N.V. Deliveries & \cellcolor[HTML]{F7C9AB}N.V. Deliveries & \cellcolor[HTML]{F7C9AB}N.V. Deliveries & \cellcolor[HTML]{F7C9AB}Newborn Sepsis & \cellcolor[HTML]{F7C9AB}N.V. Deliveries \\

\cellcolor[HTML]{F7C9AB}NCD Screening & \cellcolor[HTML]{F7C9AB}NCD Screening & \cellcolor[HTML]{F7C9AB}NCD Screening & \cellcolor[HTML]{F7C9AB}Normal Newborn & \cellcolor[HTML]{F7C9AB}NCD Screening \\
  
\cellcolor[HTML]{F7C9AB}SBCC & \cellcolor[HTML]{F7C9AB}SBCC & \cellcolor[HTML]{F7C9AB}SBCC & \cellcolor[HTML]{F7C9AB}N.V. Deliveries & \cellcolor[HTML]{F7C9AB}SBCC \\
   
\cellcolor[HTML]{F7C9AB}EPI/IMCI & \cellcolor[HTML]{F7C9AB}EPI/IMCI & \cellcolor[HTML]{F7C9AB}EPI/IMCI & \cellcolor[HTML]{F7C9AB}SBCC & \cellcolor[HTML]{F7C9AB}EPI/IMCI \\
    
\cellcolor[HTML]{F7C9AB}FP Short Acting & \cellcolor[HTML]{F7C9AB}FP Short Acting & \cellcolor[HTML]{F7C9AB}FP Short Acting & \cellcolor[HTML]{F7C9AB}EPI/IMCI & \cellcolor[HTML]{F7C9AB}FP Short Acting \\
     
\cellcolor[HTML]{F7C9AB}Growth Monitoring & \cellcolor[HTML]{F7C9AB}GM, SAM mngmt & \cellcolor[HTML]{F7C9AB}GM, SAM mngmt & \cellcolor[HTML]{F7C9AB}GM, SAM mngmt & \cellcolor[HTML]{F7C9AB}GM, SAM mngmt \\

\cellcolor[HTML]{F7C9AB}ANC/PNC & \cellcolor[HTML]{F7C9AB}ANC/PNC & \cellcolor[HTML]{F7C9AB}ANC/PNC & \cellcolor[HTML]{F7C9AB}FP all methods & \cellcolor[HTML]{F7C9AB}ANC/PNC \\
 
\cellcolor[HTML]{F7C9AB}Lim. curative care & \cellcolor[HTML]{F7C9AB}Lim. curative care & \cellcolor[HTML]{F7C9AB}Lim. curative care & \cellcolor[HTML]{F7C9AB} ANC/PNC & \cellcolor[HTML]{F7C9AB}Lim. curative care \\

\hline

\multicolumn{5}{c}{} \\
% \hline
\end{tabular}}

%%%%%%%%%%%%%%%%%%%%%%%%%%%%%%%%%

\begin{tabular}{ |l|l| } 

\hline
 \cellcolor[HTML]{F7C9AB} & Minimum Standards by facility level \\ 
 \hline
 \cellcolor[HTML]{FFFFFF} & Extra Services  \\
 \hline
\end{tabular}
%%%%%%%%%%%%%%%%%%%%%%%%%%%%%%%%%%

\caption{Minimum Standards and Extra Services by facility level which are provided by the government of the peoples republic of Bangladesh \cite{2}}
\label{table:minimumStandards}

\end{table*}

\section{Related Work}

When our proposed solution becomes operational, it will deal with highly sensitive data that must be managed in a secure manner. Since the  electronic health record (EHR)~\cite{ehr} is a part of this system, it will contain a lot of personal data, most of which is considered sensitive data. This is because , from names, addresses, ID card numbers, insurance numbers to medical history, all these data will be stored here. 
Moreover, this medical data will be updated and shared on a regular basis, provided the patient has given consent.

Almost all first-world countries operate their health services through various types of digital platforms. Currently, there are several approaches where EHR systems are enhanced with blockchain-based services. Various tech giants are also offering different cloud-based \textit{Software As A Service} (SaaS) solutions. But here \textit{privacy-protecting}, \textit{General Data Protection Regulation} (GDPR)~\cite{gdpr} and \textit{performance/scalability} are the real concern. Comparing some related work in this area, we can see that each service has had to compromise in some cases. Table \ref{table:compareTable} presents a comparison based on different features of different existing systems.
%%%%%%%%%%%%%%%%%%%%%%%%%%%%%%%%%
\begin{table*}[!htbp]
% \centering
\resizebox{\textwidth}{!}{\begin{tabular}{|c|c|c|c|c|c|} 
% \rowcolor{lightgray}  
\hline
\textbf{Reference} & \textbf{Framework} & \textbf{Type} &  \textbf{Privacy-Preserving} & \textbf{GDPR} & \textbf{Performance/Scalability} \\ 
 \hline
  \cite{roehrs2017omniphr} & Peer-to-peer & Private & \xmark & \xmark & \cmark \\
 \hline
  \cite{bocek2017blockchains} & Ethereum & Private & \xmark & \xmark & \xmark \\
 \hline
 \cite{liang2017integrating} & Hyperledger Fabric & Private & \xmark & \xmark & \cmark\\
 \hline
 \cite{al2017medibchain} & Ethereum & Private & \xmark & \xmark & \cmark\\
 \hline
 \cite{ichikawa2017tamper} & Hyperledger Fabric & Private & \xmark & \cmark & \xmark\\
 \hline
 \cite{azaria2016medrec} & Ethereum & Public & \xmark & \xmark & \xmark \\
 \hline
  \cite{nchinda2019medrec} & Ethereum & Public & \cmark & \xmark & \cmark \\
 \hline
 Proposed solution & Hyperledger Fabric & Private & \cmark & \cmark & \cmark \\
 \hline
\multicolumn{6}{c}{} \\
% \hline
\end{tabular}}

\caption{Comparison of related works.}
\label{table:compareTable}

\end{table*}
%%%%%%%%%%%%%%%%%%%%%%%%%%%%%%%%%%
The Distributed Personal Health Record System (PHR) \cite{roehrs2017omniphr} proposed by Roehrs et al. is a decentralized system that functions like a centralized system among participating devices. To describe the concept in more detail, we can use the term \textit{'peer-to-peer'}. The way the author has presented the architecture in detail, scalability can be a plus point for this system. But when it comes to the concept of PHR system, it should be controlled by the patients. But the author has also mentioned that different organizations can adopt this system for the practice. This raises concern about security and privacy. Also, in some cases, it violates the GDPR.
%%%%%%%%%%%%%%%%%%%%%%%%%%%%%%%%%%%%%%%%%%
Bocek et al.\cite{bocek2017blockchains} have proposed an Ethereum network-based  \textit{proof-of-concept} (PoC) on the pharmaceutical supply chain. However, it is mentioned here that various IoT devices can be used to communicate directly with a blockchain node server via an HTTP protocol and store this data in PostgreSQL databases. At this point, the lack of decentralization is noted. This is because if this single node's database is breached by attackers or another cyberattack occurs, data privacy may be compromised. Since there is a risk of data being exposed, most of the requirements of the GDPR are not met.
%https://www.modum.io/
%%%%%%%%%%%%%%%%%%%%%%%%%%%%%%%%%%%%%%%%%%%%%%%%%%%%%
The solution of Liang et el.~\cite{liang2017integrating} is based on Hyperledger Fabric. This has good control over data transparency and can store data in immutable ledger. But in the architecture, end user-privacy is not the main concern. While reviewing this paper, some gaps were found in the technical details. The paper makes extensive use of many wearable devices. But how these third-party components share data with their blockchain network has not been properly described. So, there are many questions left unanswered for GDPR guidelines. On the other hand, this solution was developed on Hyperledger Fabric version 0.5 where the private data collection feature was unavailable, and the authors did not propose any custom solutions. So this is a disadvantage for privacy protection. However, since the whole system is cloud-based, there is an advantage for scalability.
%%%%%%%%%%%%%%%%%%%%%%%%%%%%%%%%%%%%%%%%%%%%%%%%%%%%%%%%%%%%%%%
\textit{MediBchain}, proposed by Omar et el.~\cite{al2017medibchain}, is an Ethereum-based solution built on a cloud-based server infrastructure. This is the plus point for the scalability of the system. The cloud infrastructure does provide some scalability, but since the data of the stakeholders is encrypted, new collaborations and linking of different organizations can be a problem. Security and proper control of data transparency are required for proper authority.
%%%%%%%%%%%%%%%%%%%%%%%%%%%%%%%%%%%%%%%%%%%%%%%%%%%%%%%%%%%%%%%%
A permissioned blockchain solution that collects health data from mobile devices was proposed by Ichikawa et al.~\cite{ichikawa2017tamper}. This solution is also based on an older version of Hyperledger Fabric, which faces the same problem of private data collection. This thwarts some privacy standards. Moreover, this solution relies on the PBFT algorithm for the consensus mechanism. Thus, with Hyperledger, if an attacker manages to attack the principal of the blockchain network for more than (N-1)/3 at a time, there is a possibility that the entire blockchain service can be disabled \cite{swan2015blockchain}.
%%%%%%%%%%%%%%%%%%%%%%%%%%%%%%%%%%%%%%%%%%%%%%%%%%%%%%%%%%%%%%%%
\textit{MedRec} by Azaria et al.~\cite{azaria2016medrec} is an Ethereum-based solution for decentralized EHR management. This solution has some scalability issues. MedRec does not take care of the security of individual databases, which a local administrator must manage. It also does not solve the problem of digital rights management. Since it is an open-source project, the author would like others to contribute to its further development and help solve the current problems.
%%%%%%%%%%%%%%%%%%%%%%%%%%%%%%%%%%%%%%%%%%%%%%%%%%%%%%%%%%%%%%%%%%
The work of Nchinda et.al, \cite{nchinda2019medrec} is a follow-up to Azaria et.al, \cite{azaria2016medrec}. Both projects are from the same maintainers and have the same name (MedRec). This work \cite{nchinda2019medrec} is mainly about an EHR management system from MIT. The earlier work \cite{azaria2016medrec} is mainly about improving medical record tracking. Scalability issues are solved here. Although patients have more control over their information even if the network type is public. But \textit{proof-of-work} is used as the consensus algorithm, which has already been identified as an energy inefficient process. So the cost rate is high ,and as a service it is also dependent on third-party miners. Another problem is that personal data is stored outside the chain. Therefore, some users might have a problem with the authenticity of the data. This also violates the basic data protection regulation in some cases.
%%%%%%%%%%%%%%%%%%%%%%%%%%%%%%%%%%%%%%%%%%%%%%%%%%%%%%%%%%%%%%%%%%

In our proposed solution, in order to perform any kind of task in blockchain, it needs to execute the logic from a smart contract. In order to access it, each user needs to use a health card for authentication. These health cards contain the user’s identity. Each user can only perform the tasks they are authorized for. This way we can maintain the privacy and access of the user. European General Data Protection Regulation (GDPR) consists of seven principles described in article 5.1-2 of the general data protection regulation \cite{gdpr}. The primary scopes based on these principles are as follows: When a system automatically or manually processes data, it should be for legitimate purposes, and our proposed solution works specifically for the healthcare domain and aims to advance both healthcare research and service. Data processing must be lawful, fair, and transparent to the data subject in accordance with GDPR. As we design this blockchain-based solution, maintaining transparency based on the type of organization is the forte of this cutting-edge technology. In the case of Data Minimization and Accuracy, our system collects only a citizen's medical data, and due to its design, it dynamically updates all user data, thus adhering to the GDPR principles. The proposed system is an excellent solution for the GDPR's integrity and confidentiality principles. Because the entire system runs on a private blockchain network, its administration and governance mechanisms are well-established. All the information blocks contain encrypted data at the same time. This also meets the Healthcare Security Regulations (HIPAA) requirements for administrative safeguards \cite{badr2019blockchain}.

\section{Problem Statement}

Keeping documents and their preservation for administration is not a new thing in human civilization. For this reason, in the past, we can observe the trend of data storage from cave walls to papers, and in modern times, it is the servers in digital form. However, nowadays the main focus is on digital data because of increasing dependencies and reliability on digital platforms. When stored data is in physical form, any tampering can be detected on that copy of the data. Even on the printed surface, many security measures can be taken to detect the validity of the document, like passports and currencies for example. Both have so many factual printing patterns that they can be easily distinguished from other counterfeit copies. But in case of digital data, this is where the biggest challenge lies. Because they can be altered or deleted without a trace. If an intruder can manipulate the data from the central database, this application will print a new copy with fake data for other users. In this way, any national/international threat can arise.

In Bangladesh, the government has a precise administrative structure and chain of command for maintaining public medical care and health services. The central administration also allocates an estimated budget every year for maintaining this system. But like all other sectors, this sector also has its downsides. Sometimes there are various allegations against some of the system's officials, leading to public suffering in the medical industry. The department has its own quality assurance policy. But since the department is so big and various organizations are working here. Many consumers cannot even approach the higher authority with their complaints. There is a massive communication gap between the higher authority and the consumers. Most consumers do not even know that the services they want are among their rights already granted by the department. Nevertheless, when the authority receives a complaint, sometimes no legal action can be taken because proper documents and evidence are not available. As a result, some unethical individuals take advantage of these benefits to earn more money in an unethical manner. Following  problems are identified by observing the scenarios.
\begin{itemize} 
\item Doctors influenced by private drug  distributors. 
\item Medications provided for free by the government are being sold. \item Medical records are not properly kept. In addition, the records that are kept are difficult to find. 
\item The procedures for feedback and complaints are very critical. \item As the medical staff does not care about the quality assurance of the services, the consumers often have to suffer a lot in the hospitals. 
\item Due to the lack of monitoring, some unauthorized doctors take advantage of the benefits. 
\item Anyone can easily falsify their qualification status to increase their value. 
\item Most patients have no knowledge of their own medical records. As a result, doctors cannot properly investigate their illnesses. 
\item Most importantly, the consumer's interaction with the service provider and its authority is very unpleasant, so most people switch to private services and those who cannot afford it have to endure countless suffering in the public service.
\end{itemize}

\section{Proposed Solution}

The government is already familiar with keeping and of working with ledgers in paper form. Distributed Ledger Technology (DLT)~\cite{dlt} can be a great solution that can easily be adopted in this sector as a digital service. Conventional technology may be enough to do normal office work. But we have described the scale of this sector and other problems in above sections. It is clear that this sector requires much smarter transparency, correctness, validity, authenticity, and reliability. Most importantly, this service needs to be more secure at the national level to prevent any kind of information manipulation. For this reason, it can be claimed that this sector should be introduced with a brand new technology commonly known as blockchain~\cite{miah_bc}, a form of DLT. This technology can be recommended because blockchain has gained the most attention among other new solutions in the industry, government and academia \cite{5}. It has some specific development standards and strict set of rules by which it can ensure its security from the beginning ,and after its functional operation it generates and maintains a series of blocks by which it can form a ledger with these successive blocks \cite{5}.
\newline
Although our paper contains some examples on financial topics and blockchain has been introduced in our proposed solution. We would like to clarify that our proposed solution does not cover the service of a payment system and any kind of cryptocurrency~\cite{crypto} issues are not discussed. Our main concern in this research project is to propose a digital information management system that ensures public health service status, information transparency and security.

% \subsection{Why our Solution?}

The solution we have proposed corresponds directly to the national demand in Bangladesh. More specifically, the government of Bangladesh has already published the "National Blockchain Strategy" in which the need of the health application area has been specified \cite{6}. There, the limitations of the current system have been mentioned by the government itself \cite{6}. The importance and demand for a privacy-friendly system is cited as a perfect solution in this area. The collection, storage and retrieval of health data while maintaining privacy is the key point of the demand raised by the government \cite{6}. In our solution, we provide the same elements and in our prototype, these features are also functional with the blockchain solution. Most importantly, Bangladesh has a well specified and strict law under the "Bangladesh Digital Security Act" due to which the government of Bangladesh cannot subscribe to public cloud services of blockchain. Though these public cloud services of blockchain are easy to manage, but it is not suitable to store sensitive national data in such third-party services. For this reason, Bangladesh is already convinced of the importance of a "National blockchain Platform" hosted by the relevant agency of the Government of Bangladesh \cite{6}. So, we have developed a prototype as a Proof of Concept (PoC) employing a reputed framework "Hyperledger Fabric", hosted on our own server and globally redirected through a domain of our own DNS server~\cite{dns} and followed by the use case Health Application Domain. We anticipate that our proposed solution can be a potential step towards the adoption of blockchain technology for the nation. As our proposed system can be set up on our own local server, and it can be hosted globally with a domain name. So this system can be set up in the national data center.

% \subsection{Accelerate Clinical/Biomedical Research}

\begin{figure}[!htbp]
    \includegraphics[width =\linewidth]{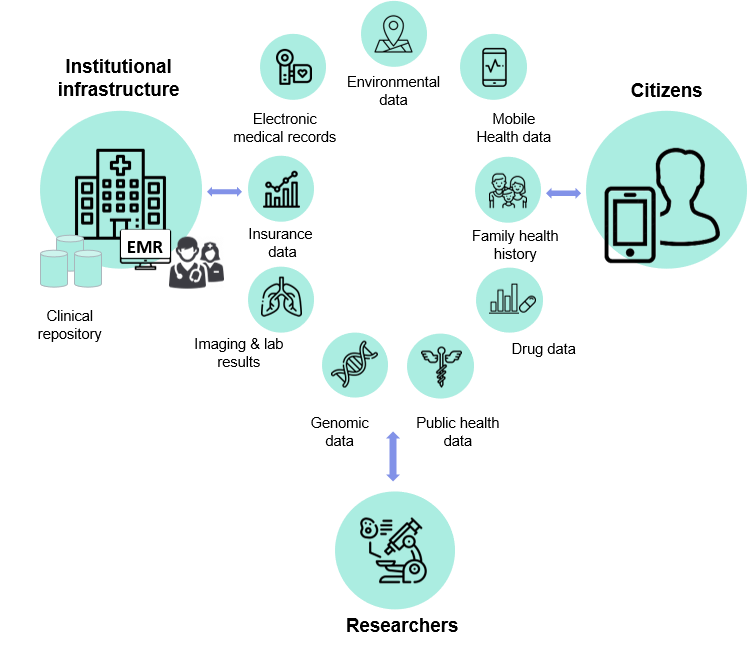}
    \caption{The key component types from Electronic Health Record (EHR) Service which can be kept inside a secure Blockchain Network to exhilarate the research in Medical Science and Computer Science}
    \label{fig:ehrCycle}
\end{figure}

With several government and non-government organizations involved in clinical / biomedical research, a decentralized blockchain service may be the best choice in terms of secure and traceable collaboration with resource management features. With the help of this technology, various organizations can collaborate with other institutions/organizations to share and analyze data without relinquishing control. Each institution can maintain full control over its own computing resources while collaborating with other institutions. In this way, a solid data set  can be formed for each clinical trial \cite{badr2019blockchain}. Fig \ref{fig:ehrCycle} represents the overview of this idea of secure collaboration.

In this way, patient-generated data becomes available to researchers. As a result, a decent environment as shown in Fig \ref{fig:ehrCycle} can be created where academia, healthcare industry and healthcare professionals can collaborate.
 
In case of security and privacy, private blockchain can be a good option to control transparency, security and immutability of data. If anonymized and then tracked in the research process with a timestamp, this secondary data source would enable millions of individuals, healthcare providers, healthcare institutions, and medical researchers to share vast amounts of genetic, nutritional, lifestyle, environmental, and health data with guaranteed security and privacy \cite{badr2019blockchain}

\section{System Design}
Private blockchain network is suitable to develop such robust system. This is also known as Permissioned Ledger Technology \cite{5}. To demonstrate the proof of concept of the proposed solution, a prototype is developed with the renowned blockchain framework “Hyperledger Fabric” which is well suited to the scenario of the proposed solution ~\cite{cbfha2019Wiley}. 
\subsection{Why Private Ledger?}
% Our proposed project is one kind of management system which will cover the services of 
% administration, health official's information, authorized drug / medicine list and patient's medical history. So there is a need of preservation of all these data. For that privacy in here is very important. Specially when this system deals with the data of a government officials, data transparency control is a big challenge. But if we try to solve this issue with private blockchain network the service can easily maintained. Because here every user will be well identified and will get enrolled by specific administration, will belong under an exact organization. So no anonymous or unknown user or bad actors cannot get in the network to perform any attack in the system. On the other hand Public Ledger is one kind of open network where anyone can join or leave. The transparency is very 
% clear to everyone. So the privacy concern causes for some of the scenario like this project \cite{5}. On the other hand most of the Public Ledger Technology is third party dependent. Such as Miner. So for organizations like Government it is not a good practice to rely on third party dependencies. Here services and information will be transparent to everyone but privacy-preserving priority will be at maximum level. 
The proposed system is a kind of administrative system that includes the services of the administration, the information of the health officials, the list of approved drugs, and the medical history of the patients. So there is a need to keep all these data. For this reason, data protection is very important here. Especially when this system is working with the data of government officials, controlling the data transparency is a big challenge. However, with  a private blockchain network, the service can be maintained easily and the stated data transparency and control related problems can be solved without any complex procedure. Because here every user is identified precisely. Each user is registered by a particular administration and belongs to a particular organization. So, no anonymous or unknown user or bad actor can enter the network to make an attack on the system. On the other hand, public ledger is a kind of open network that anyone can join or leave. The transparency is very clear to everyone. Therefore, there are privacy concerns in some scenarios like this system mention in this paper \cite{5}. On the other hand, most public ledger technologies depend on third party vendors such as miners. So for organizations like the government, relying on third party dependencies is not a good practice. Here, the services and information are transparent to everyone, while maintaining the privacy with top priority.
\subsection{Why Hyperledger Fabric?}
% Compared to the other frameworks, Hyperledger Fabric is a completely permissioned blockchain network that is well designed for operations involving sensitive and confidential data and also the most complete solution for developing healthcare applications \cite{cbfha2019Wiley}. It was established under the Linux Foundation and also received 
% contributions from IBM, Intel and SAP Ariba to support the collaborative development of blockchain-based DLT. It has been designed with enterprise grade permissioned distributed ledger technology (DLT) to ensure that different governmental agencies and business organizations can take advantage of a DLT system in different use-cases \cite{5}. Most specifically this Permissioned Ledger technology is greatly known as immune ledger technology \cite{7}.

% So after explaining the pipeline approach of our technical plans the most common question may arise for the blockchain technology. “Are we making the government’s operational process dependent on third party Miners?” To explain this matter the term Consensus Mechanism will come up as we have mentioned that we will be following Hyperledger Fabric Framework to develop this system. \textit{This framework also hold a record of 3000 transactions per second So the traditional notion of blockchain's slow speed and long processing time will now be dispelled.\cite{cbfha2019Wiley}.}

Compared to other frameworks, Hyperledger Fabric is a fully permissioned blockchain network that is well suited for operations involving sensitive and confidential data and is also the most comprehensive solution for healthcare applications \cite{cbfha2019Wiley}. It was founded under the Linux Foundation and also received contributions from IBM, Intel and SAP Ariba to support the joint development of a blockchain-based DLT. It was developed with an enterprise-grade permissioned distributed ledger technology (DLT) to ensure that various government agencies and business organizations can take advantage of a DLT system in various use cases \cite{5}. This permissioned ledger technology is primarily known as immune ledger technology \cite{7}.
So, after explaining the pipeline approach of our technical plans, the most common question about blockchain technology is. "Are we making the government's operational process dependent on third-party miners?" To explain this, the term consensus mechanism comes up as we mentioned that proposed system utilized the Hyperledger Fabric framework. This framework also holds a record of 3000 transactions per second, so the traditional notion of blockchain's slow speed and long processing time will now be dispelled~\cite{cbfha2019Wiley}.

\subsection{Consensus Mechanism}

% Though we can see the vast use of the Mining concept in blockchain (Public Network 
% mostly) technology where Miners play the role of validating transactions. But in our case as we are developing Private blockchain Network with Fabric, we have a well proved alternative concept. That is “Ordering Peer”. As government organization itself is a national authority, this concept can fit in our system properly. Because ordering peers are not random peers in the network. These peers will be constantly selected for executing the Consensus Mechanism to validate every transaction neutrally. These peers can be one or multiple in a single channel. So the ministry's IT division can hold these peers easily at their personal datacenter and if needed can scale multiple Ordering peers easily with the help of Virtual Machines or any other virtual containers like Docker. This is how the 
% Ministry can stay self-reliant which it likes to be and also the cost will be efficient. (Figure: \ref{fig:generalArchitecture}, port:7050 and 8050 )

While we can see that the mining concept is widely used in blockchain technology (mostly a public network), where miners play the role of validating transactions. But in our case, since we are developing a private blockchain network with Fabric, having a proven alternative concept. This is "Ordering Peer." Since the government organization itself is a national authority, this concept can fit well into our system. This is because the ordering peers are not random peers in the network. These peers are constantly selected to execute the consensus mechanism to neutrally validate each transaction. These peers can be one or more in a single channel. In our system, the ministry can be the direct owner and authority of the system. From the system user perspective, the representative of the ministry can operate as governance. And as this system is developed with Hyperledger Fabric the concept of ‘Peers’ is a fundamental element of a private blockchain network. Here each peer can represent an organization that is responsible to host its related copies of ledgers and smart contracts. As Hyperledger Fabric forms its network with multiple peers that is where Docker comes in this topic. To host all these peers as a service Docker is a very good option for OS-level virtualization. These docker containers can reduce the cost of the Blockchain Architecture required infrastructures and make the whole system more manageable, scalable, and efficient. In these containers, different services are hosted from a dedicated port. 
%Thus, the ministry's IT department can easily keep these peers in its personal data center and scale multiple ordering peers as needed using virtual machines or other virtual containers such as Docker. This way, the ministry can remain independent, which it likes to be and also the cost will be efficient. 
% Port:7050 and port 8050 in Fig \ref{fig:generalArchitecture} represent the described scenario.

%  ####################  need editing of the answer

\subsection{Consensus Algorithm}

% As we have cleared the validation responsibilities in our system now it is important to know how the 
% validation will occur. For that we will be using Raft which is a distributed consensus algorithm. Because this 
% algorithm not only validates the transactions, it is also well known for its fault tolerance and fast exception handling. 
% If our network has multiple ordering peers Raft will elect one of it as Leader Peer. If the leader peer fails to respond 
% in time Raft will elect a new ordering peer as a new leader to validate the transaction.
Now that we have clarified the responsibilities for validation in our system, it is important to know how validation is done. For this Raft~\cite{raft} is used, which is a distributed consensus algorithm. This algorithm not only validates transactions, but is also known for its fault tolerance and fast exception handling. If there are multiple ordering peers in the network, Raft elects one of them as the leader peer. If the leader peer does not respond in time, Raft elects a new ordering peer as the new leader to validate the transaction.

\subsection{Data Storage}
% As our whole system is based on distributed ledger technology we 
% have maintained a special mapping for data storing. At Figure: \ref{sfig:peerstructure} and \ref{sfig:orderedstructure} the existence 
% of Block is shown, which actually holds the data. These blocks hold transactional 
% data only which are in actual bytes (1024 is 1KB). These blocks are 1 megabyte 
% (MB) in size which also holds every committed change and the Genesis block only 
% holds the hash and signatures or keys \cite{9}. These data are known as On-Chain data 
% which are at maximum level of security.
% \newline
% But our system also has to deal with large amounts of data. Such as, user's image, doctor's certificate (PDF file) and 
% other scanned documents. These types of non-transactional data are known as Off-Chain data (figure \ref{fig:offChain}). We cannot keep this 
% data in the Block for efficiency. Though base64 could be a solution. But as medical applications may run 24/7, this 
% can also be a threat for load-balancing. So for our prototype we are keeping our Off-Chain data at our Node.js based 
% web application (Centralized). But at the production level we can manage these types of files with the \textit{ \textbf{Inter Planetary File System}} 
%  (IPFS) \cite{10}. While transactions are collected inside the clocks we are using CouchDB 
% as our World State which holds current values of a ledger. This helps the system to invoke chain code/ smart 
% contract with API to execute get, put, and delete requests in a faster way for better and faster user experience.

Since our entire system is based on distributed ledger technology, we created a special mapping for data storage. In Fig: \ref{sfig:peerstructure} and \ref{sfig:orderedstructure}, we show the existence of blocks where the data is stored. These blocks contain only transaction data specified in actual bytes (1024 is 1KB). These blocks have a size of 1 megabyte (MB), in which any transmitted change is also stored, and the genesis block contains only the hash and signatures or keys \cite{9}. This data is called on-chain data, which has the highest level of security.

\begin{figure}[!htbp]
    \includegraphics[width =\linewidth]{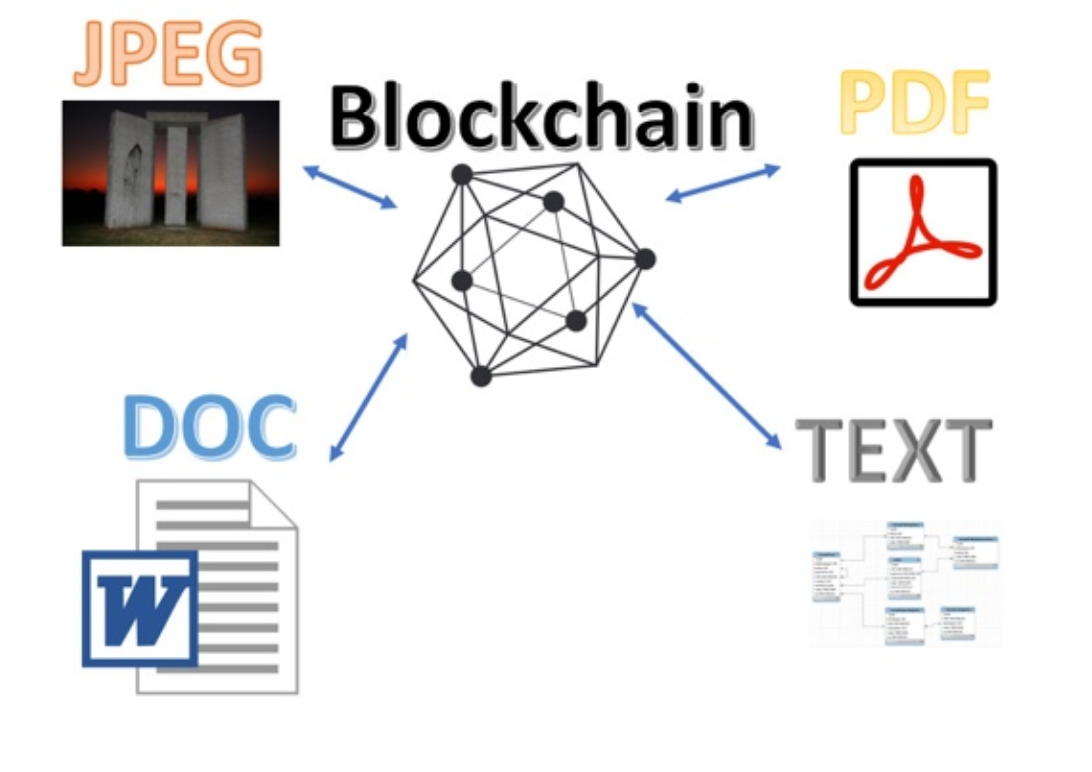}
    \caption{Example of Off-Chain data \cite{9}}
    \label{fig:offChain}
\end{figure}

In addition to this, our system also has to deal with large amount of data. For example, user's picture, doctor's certificate (PDF files) and other scanned documents. These types of non-transactional data are called off-chain data. Fig \ref{fig:offChain} represents the off-chain data. We cannot keep this data in block for efficiency reasons. However, since medical applications can run 24/7, this can also pose a threat to load balancing. So for our prototype, we keep our off-chain data in our Node.js-based centralised web application. But at the production level,  these off-chain data can be managed with Inter Planetary File System (IPFS) \cite{10}. While the transactions are collected inside the clocks, we use CouchDB~\cite{couchdb} as our World State, which contains the current values of a ledger. This helps the system to invoke chain code/smart contracts with API to execute get, put, and delete requests faster for better and faster user experience.

%In addition to this, our system also has to deal with large amount of data. For example, user's picture, doctor's certificate (PDF files) and other scanned documents. These types of non-transactional data are called off-chain data. Fig \ref{fig:offChain} represents the off-chain data. We cannot keep this data in block for efficiency reasons. Even though base64 could be a solution. However, since medical applications can run 24/7, this can also pose a threat to load balancing. So for our prototype, we keep our off-chain data in our Node.js-based centralised web application. But at the production level,  these off-chain data can be managed with Inter Planetary File System (IPFS) \cite{10}. While the transactions are collected inside the clocks, we use CouchDB~\cite{couchdb} as our World State, which contains the current values of a ledger. This helps the system to invoke chain code/smart contracts with API to execute get, put, and delete requests faster for better and faster user experience.%

% \Figure[t!](topskip=0pt, botskip=0pt, midskip=0pt)[width=.9\linewidth]{figures/OffChain_Example.png}
% {Example of Off-Chain \cite{10}.\label{fig:offChain}}

\subsection{Smart Contract}

A smart contract, also known as chain code, is an executable script that is triggered when it meets its conditions~\cite{miah_bc}. This is the file that contains the business logic of the system.  It runs as a script during the execution phase \cite{7}.This also controls the capability of information read and write. More specifically, this file allows to write any logic and set conditions in the computer code according to the system requirements and specifications. In Hyperledger Fabric, various SDKs are provided that allows to write chain code or smart contract in JavaScript. This chain code can have different logic and conditions depending on the role of the organization and each supporting peer of that organization has the same chain code and it is depicted in the Fig \ref{sfig:peerstructure}. When a request appears to these peers, the chain code checks the condition of this request and if the conditions are met, the execution will make a change in the world state (DB) and respond with a signature. Steps 2 and 3 of Fig \ref{fig:favricGovernance} references to this operation. In Fig: \ref{fig:chainCodeMapping} a brief overview of the chain code or smart contract is depicted.
\newline
Our blockchain service is fully functional as a backend service and the smart contract plays an important role here. Many autonomous triggers are managed here for better user experience and faster service. Since chain code/smart contract programming in Hyperledger Fabric provides support for various programming languages like Java, Go and JavaScript \cite{cbfha2019Wiley} we used the JavaScript SDK to write smart contracts. We also created a web portal using the Express.js framework~\cite{expressjs}, which provides various APIs that can be used to communicate with other medical or IoT devices. The proposed system is flexible and scalable. So in the future, if the healthcare sector adopts IoT devices on a large scale and global technology provides so much transparency in terms of privacy, this sensor data can be easily captured by creating more APIs and methods in our smart contracts / chain codes.

% \textit{Smart Contract which is also known as Chaincode in technical term. This is a file which holds the general data, business logic and its preserving architecture. It runs as a script during the execution phase \cite{7}.This also controls the capability of information read and write. So when a Chaincode gets deployed it gets installed in all the Endorsing Peers (Figure: \ref{fig:favricGovernance} and \ref{sfig:peerstructure}) of the DLT. So the Chaincode is also distributed in DLT.
% }

\subsection{Stakeholders}
% Our solution as a product will be available in three different platforms. There will 
% be Web portal, Computer software and Mobile application. These platforms are separated by the user types. Our users 
% are: 
Our solution as a product will be available on three different platforms. There will be a web portal, desktop application and a mobile application. These platforms are divided by user type. Our users are:
\begin{itemize}
  \item Central Authority (In the prototype, BMDC is playing this role.)
  \item Doctors
  \item Citizen/ Patient (In our prototype this type of user is known as “Nagorik”)
\end{itemize}

\subsubsection{Central Authority}
Among these three types of user The Central Authority can be any organization or ministry of the central government. For example, “Ministry of Health and Family Welfare”. This type of user can perform the following tasks from the web portal.
\begin{itemize}
  \item Authorize Doctors, Medicines.
  \item Analyze Health data with privacy (Anonymous).
  \item Review complains submitted by the patients.
  \item Can verify and approve the update of any doctor's information.
  \item Can check government authorized medicine / drug list.
  \item Can store data of medicine distribution for future investigation.
  \item Can store data to analyze Doctors' drug prescribing tendency for the future.
  \item See official news feed for every verified health news to keep the user updated.
\end{itemize}

\subsubsection{Doctors}
% The Doctors have to be authorized by the Authority User of this system. Authorized doctors will have a desktop software. This Software can operate online/offline in both situations. With this software our system can easily preserve every data and for further assistance it can also help the doctor for the patient's case study. As it can keep track of every data, the activity of every doctor who are working as a health service provider can be analyzed. In general operation The Doctor's software can:

The Doctors must be authorized by the central authority user of this system. Authorized physicians will be provided with desktop application. This software can work both online and offline. With this software, the system can keep all the data to help the doctor in case study of the patients. As the software keeps all the data, the activity of any doctor who is a health care provider can also be analyzed. In general operation, using the software the doctors can perform the following tasks.

\begin{itemize}
  \item Create / print prescription and automatic assistance for suggesting proper Medicine.
  \item Alert notification if any drug/medicine can be a threat for current patient.
  \item Store every general information efficiently to assist the doctor during poor network connection.
  \item Get fast identification of the patient through his/her NID card or Birth Certificate.
  \item Doctor can update their educational / experience status from the software and after verifying the validity of that information it will get published.
\end{itemize}

\subsubsection{Nagorik}
% The Patient or the user type known as “Nagorik” can be easily identified with the national identity card number. So each identified user will have a Web portal to have national healthcare services online. The services are:

Nagorik user are the general people of Bangladesh. They are the patients who will seek medical services from our system. They can be easily identified by the national identity number. Each identified user has an account in the web portal through which they can access national health services online. The services are:
\begin{itemize}
  \item Patient / citizen can login and logout to his / her own “Nagorik Portal”.
  \item Can see all his / her medical records (Medical Test/ Prescription) online with full privacy. 
  \item Can check the authorized medicine / drug list and also about the free medicines and tests.
  \item Can place complain easily if he / she is not satisfied with any government medical service.
  \item Can check his complain priority status whenever he / she wants.
  \item Can find a specific Specialist Doctor according to patient's problem. ( Verified Specialty )
  \item Can get easy appointment from home. So the doctor and patient both can maintain a well timing. 
\end{itemize}

% \textit{So as we can see this proposed solution is covering most of the requirements of Healthcare and Medical services. It
% can collect every single data related with this service. So whenever needed this data can be used as an evidence to 
% investigate any type of complains against any service provider. So for the neutral data preserving all the 
% stakeholders can stay secure if he / she is honest with his / her job duty. Any user cannot take advantage of this system 
% to bring false allegation against any service provider.}

So, as we can see, this proposed solution covers most of the needs of healthcare and medical services. It can collect all the data related to this service. So, if needed, this data can be used as evidence to investigate complaints against a service provider. By keeping the data neutral, all parties involved can be sure when they are fulfilling their duties. No user can exploit this system to make false accusations against a service provider.

\subsection{Architecture}

\begin{figure}[!htbp]
    \includegraphics[width =\linewidth]{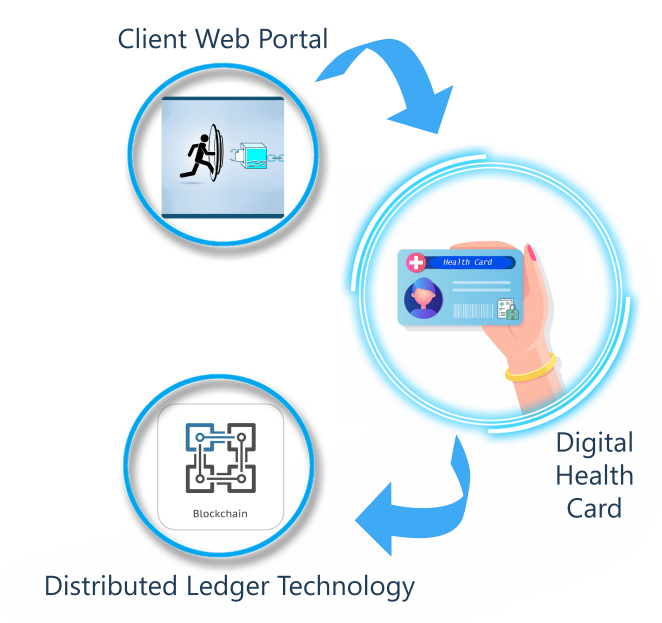}
    \caption{The Bridge of User and DLT where user can gain access in the Blockchain environment with his / her dedicated health-card which will be stored at SDK based application server.}
    \label{fig:userNdlt}
\end{figure}

If we generalize the architecture of our proposed blockchain application, we can simply split it into two parts. The "SDK" part and the "Decentralize Ledger Technology". The "SDK" part is the point that connects the user directly to the DLT solution. The availability of developer resources for Node.js and the "Hyperledger Fabric" framework is sufficient, so we prefer the "Express JS" framework in the "SDK" part. As these two frameworks and languages are popular in Bangladesh, a well-maintained interactive service can be developed to connect people with this new technology. So we get the flexibility of rapid development to enable the "Human Computer Interaction" (HCI). Fig \ref{fig:userNdlt} represents the architecture in three parts to illustrate the approach.

\begin{figure*}[!t]
    \centering
    %\vspace*{-.02in}
    \subfloat[A detail overview of the system where different types of user will be operating with their dedicated user platform to perform their operations in the Blockchain environment]{
        \hspace*{-.1in}
        \includegraphics[width=\linewidth]{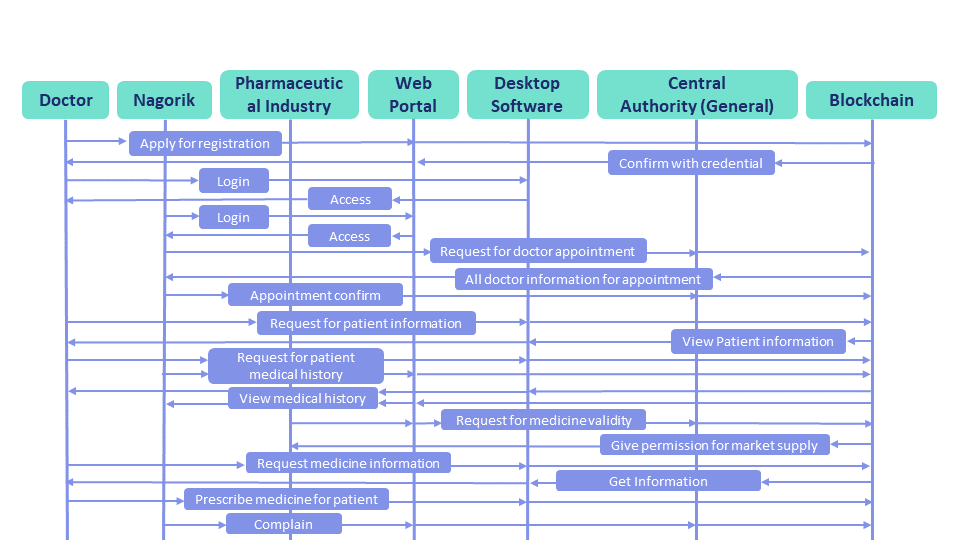}
        \label{sfig:systemArch}
    }
    %   \vspace*{-.14in}
    \hfill
    \subfloat[A sequence diagram of Doctor's Desktop application communicating with three different types of peers of Blockchain Network through SDK based application]{
        \hspace*{-.1in}
        \includegraphics[width=\linewidth]{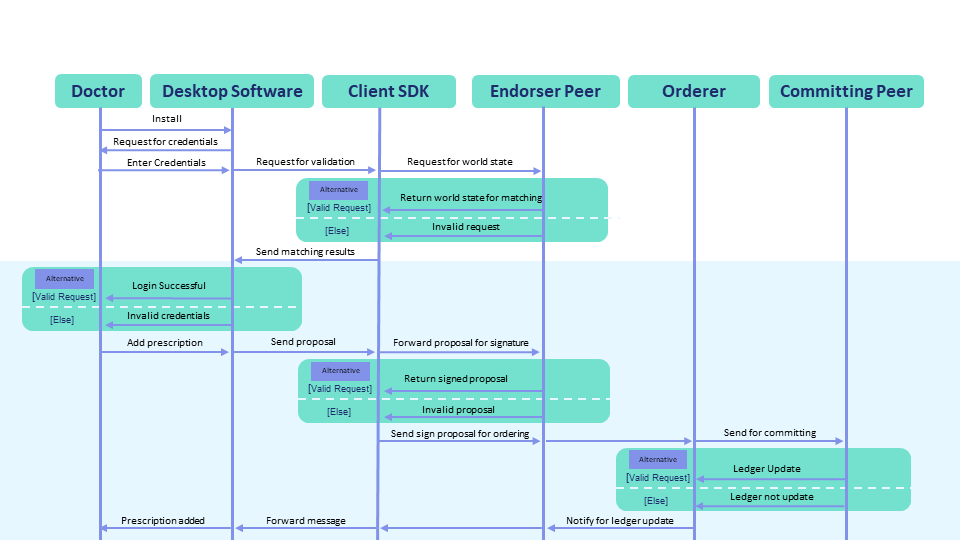}
        \label{sfig:dataFlow}
    }
    %   \vspace*{-.14in}
    \caption{Platform overview and Doctor's application sequence diagram}
    
    \label{fig:systemArchDataFlow}
\end{figure*}

% Our system's architecture currently have three user type which are Doctor, Nagorik and Central Authority. And two platforms from where services are given Web Portal and Desktop. 

% In \ref{sfig:dataFlow}, data flow of a simple process can be seen. Here doctor first installs the Desktop Software and the software will ask for credentials. Then doctor will enter their credentials then this request will be sent to the Endorser Peer by Client SDK. If the request is invalid, the invalid request message will be sent back to Client SDK and if its valid the world state for matching will be sent back to the Client SDK. After that the matching results will be sent to the Desktop Software. If the result matched then the doctor will be logged into the system else the request will be declined.
In Fig \ref{sfig:systemArch} we can see a sequence diagram of our system. First the doctor applies for registration in order to use the system. The request will be sent to the Blockchain and the Central Authority will validate the registration and sent back the confirmation to the Doctor. After that the doctor can login into the system if given access by the Desktop Software. 
Nagorik can log into the system if given access my web portal. After successfully logging in a Nagorik can request for doctor appointment. The request will be sent to Blockchain and it will sent back the doctor appointment data back to requester. Then Nogorik can confirm the appointment and this will be updated at Blockchain.
Doctor can request for patient's information and medical history in Desktop Software if given consent by the patient. The information will be taken from Blockchain and will be sent back to doctor through Desktop Software. Doctor can request for medicine validity. The request will be sent to Blockchain and if its valid then permission will be given for market supply to the doctor. Doctor can prescribe medicine for patient. The prescription information will be sent to Blockchain and the Blockchain will be updated.
Nagorik can also complain using Web Portal and the complains will be sent to Blockchain and if valid, they will be added to the Blockchain. 

The architecture of the proposed system currently has three user types, namely doctor, nagorik and central authority. And two platforms from where services are provided: Web portal and Desktop application. In Fig \ref{sfig:dataFlow}, the data flow of a simple process can be seen. Here, the doctor first installs the desktop software and the software asks for credentials. Then the doctor enters his credentials and this request is sent by the client SDK to the endorser peer. If the request is invalid, the invalid request message is sent back to the client SDK, and if it is valid, the world state is sent back to the client SDK for matching. After that, the matching result is sent to the desktop software. If the result matches, the doctor is logged into the system, otherwise the request is rejected.

\begin{figure*}[!htbp]
    \includegraphics[width =\linewidth]{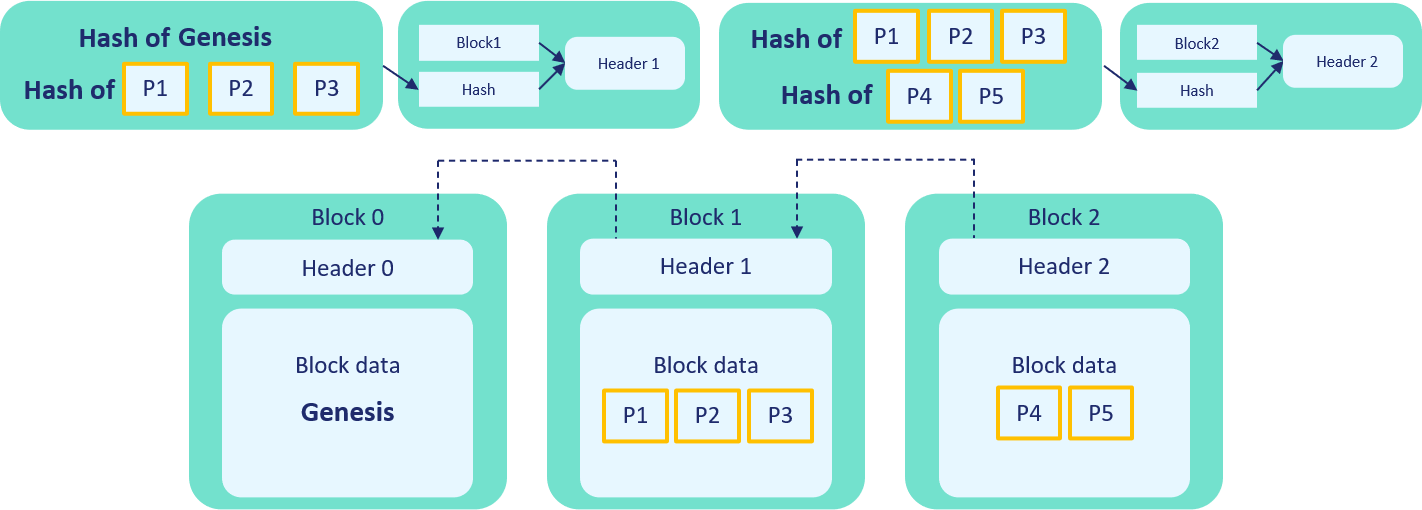}
    \caption{Generic Blockchain Architecture~\cite{miah2020introduction}}
    \label{fig:blockArch}
\end{figure*}

% As we can see in \ref{fig:blockArch}, at the beginning there will be a genesis block. And after passing all the endorsing and governing process a new block will be added to the chain. The current block's hash will be added with the newly added data the value will be merged with a new block number and new block's header will be created. For the next new block the process will be same. From this we can see that every block actually related with a chain of sequence in the blockchain. So from any point of the chain any information can not be manipulated. That is why blockchain defined as an immutable ledger.

As we can see in the Fig \ref{fig:blockArch}, there is a Genesis block at the beginning. And after it goes through all the approval and governing processes, a new block is added to the chain. The hash value of the current block is added with the newly added data, the value is merged with a new block number, and the header of the new block is created. The same process is performed for the next new block. From this we can see that each block is actually linked to a chain of sequences in the blockchain. So, no information can be manipulated from any point of the chain. For this reason, blockchain is defined as an immutable ledger.

\begin{figure}[!htbp]
    \includegraphics[width =\linewidth]{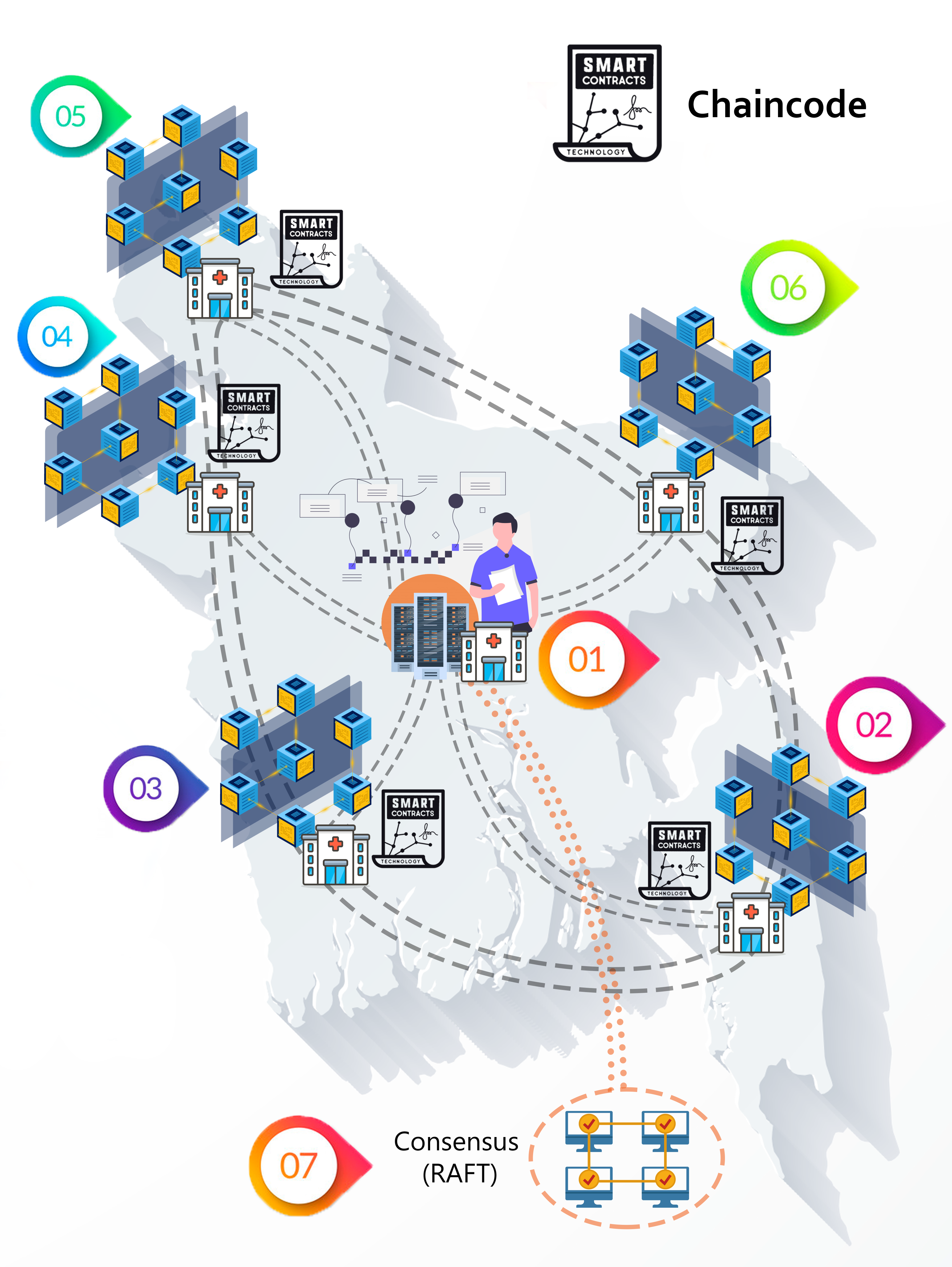}
    \caption{A visual representation of national scale Blockchain service where every public / private medical institutions are acting as Blockchain Nodes}
    \label{fig:chainCodeMapping}
\end{figure}
%%%%Figure: \ref{fig:chainCodeMapping}

\subsubsection{Client Web Portal}
% A public website developed with traditional web technology and will be identified by an open domain name so that anyone can easily find and visit this. For general visitor this website's front-end will act like a regular web page. But for the people of Bangladesh who holds a NID, this website's back-end will have a mechanism to open a portal to have national online health service which will run and secured with DLT technology.
A public website developed using conventional web technology and identified by a public domain name so that anyone can easily find it on the internet. For general visitors, the front end of this website behaves like a normal website. But for the people of Bangladesh with national identification number (NID), the backend of this website contains a mechanism to open a portal for a national online health service secured with DLT technology. This portal is developed with Node.js and Express.js Framework.

\subsubsection{Digital Health Card}
% As our proposed solution is a private DLT. So every user / citizen cannot join or leave without proper authorization. So every in the DLT end there will be a dedicated organization who will verify each citizen with the NID and an admin of that organization will enroll them as user for once in that organization, so that they access in the DLT services. This enrollment process will includes the Digital Signature of that organization and generate a digital health card for that specific user. (The concept of this health card is relatable with the concept of wallet of other private DLT.)
The solution we propose is a private DLT. So, any user/citizen cannot join or leave without proper permission. Therefore, at the  the DLT end, there will be a special organization that will verify each citizen with the NID and an administrator of this organization will register them once as user in this organization so that users can access the DLT services. This registration process includes the digital signature of this organization and creates a digital health card for registered users. This health card is similar to the wallet concept of other private DLTs.

\subsubsection{DLT - blockchain as a Service}
% After successful enrollment when a registered 
% user sends a request trough the SDK portal 
% with the Health card. Here every request from 
% the user must maintain a proper Protocol and a 
% specific structure with many authentic 
% parameters. Every request must verify the 
% user first and then if the request is valid 
% according to the Smart Contract (Figure: \ref{fig:chainCodeMapping})
% then the request gets executed and sends a 
% response to the SDK. After every successful
% execution the information gets archived with 
% its present and past from in all its distributed 
% ledgers. This is how information is preserved 
% in a sequential form of blocks, which 
% represents the term blockchain.
After successful registration, a registered user sends a request through the SDK portal using the health card. Here, each request from the user maintains a proper protocol and a specific structure with many authentic parameters. For each request, first the user is verified and then if the request is valid as per the smart contract then the request is executed and sends a response to the SDK. After each successful execution, the information is archived with its present and past form in all distributed ledgers. In this way, the information is kept in a sequential form of blocks, which represents the term blockchain. Fig \ref{fig:chainCodeMapping} represents the overview of this operation.

\begin{figure*}[!htbp]
    \includegraphics[width =\linewidth]{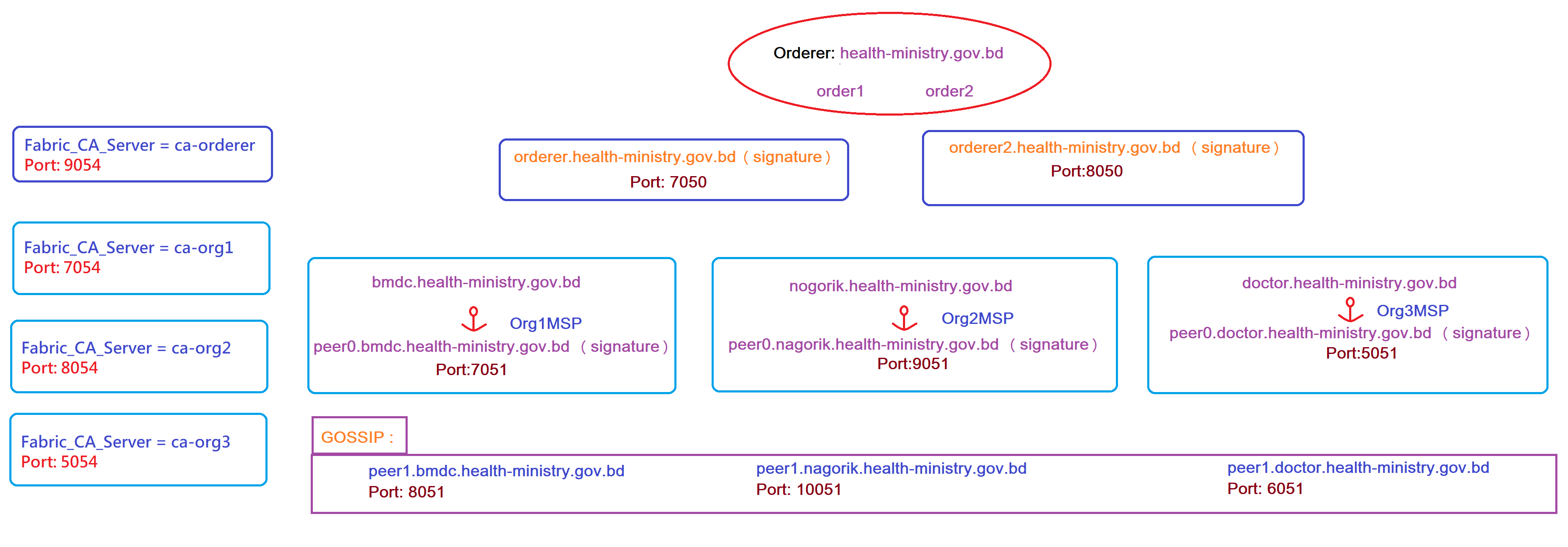}
    \caption{Base servers of the system which will run on different ports at Docker Containers}
    \label{fig:generalArchitecture}
\end{figure*}

\begin{figure}[!htbp]
\centering
    \includegraphics[width =.7\linewidth]{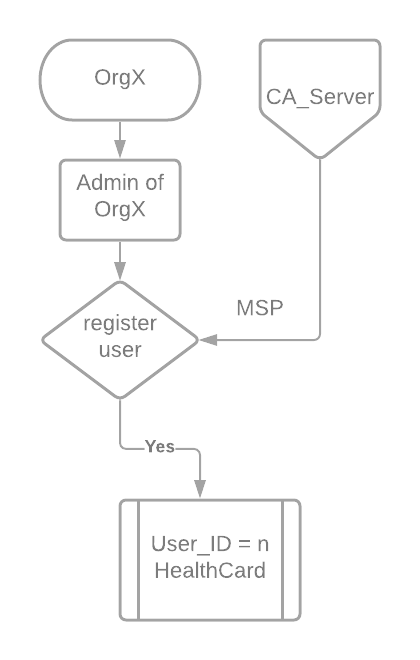}
    \caption{User Enrollment process presented as a flowchart}
    \label{fig:addUser}
\end{figure}

The proposed solution has a well-structured chain of command to maintain its governance. In the Fig: \ref{fig:generalArchitecture}, the basic infrastructure is shown as an example. Fig \ref{fig:generalArchitecture}, is considered as a virtual machine environment. Here, it can be seen that multiple peers hold different ports of a system. There is an "orderer" with two nodes (port: 7050, 8050) that run the consensus mechanism (RAFT) to generate a valid block in the blockchain sequence. In the Fig \ref{fig:generalArchitecture}, there are three different organizations (port: 7051, 9051, and 5051) that will govern the individual  user activity of the system. Each user type can have its own organization for its governance. These three organizations, the first peer (Port: 7051, 9051 and 5051), are called the "Anchor Peer". This is the address to which all user/client requests are sent. It holds the chain code and executes it on request~\cite{7}. The other peers can be used as a 'Gossip Peer'  holding the ports 8051, 10051, 6051 to verify the integrity of the system node through peer-to-peer block gossip~\cite{7}. These three Anchor Peered organizations each have their own administrator responsible for registering new users. These administrators are able to add new users under their organization in the network, for which the system creates a health card for each registered user. Fig \ref{fig:addUser} represents the user adding procedure to the network. Only these registered users can enter the DLT service and perform their actions in the organizations they belong to. All the actions of these users will be verified each time with their health card, and if the verification is successful, the ordering peer will verify it with its consensus mechanism. This is the basic structure of our governance property and all these organizations are identified by MSP.

% \break
%%%Fig 5

% MSP stands for “Membership Service Provider” is a service which maintains its operations from the help of 'CA Server'. In the Figure: \ref{fig:generalArchitecture} there are some nodes which hold the ports of 9054, 7054, 8054, 5054. Among these ports we can see there is a dedicated CA server for each of the organizations. Even the Orderer has its own dedicated CA server. All these servers can run the identification operation maintained by the MSP. Tools for key management and registration of nodes are also part of the MSP \cite{7}.

MSP stands for "Membership Service Provider" and is a service that maintains its operation using 'CA Servers'. In the Fig: \ref{fig:generalArchitecture} there are some nodes which contain ports 9054, 7054, 8054, 5054. Among these ports, we see that there is a dedicated CA server for each of the organizations. The orderer also has its own CA server. All these servers can perform the identification process which is managed by the MSP. Tools for key management and node registration are also part of the MSP \cite{7}.
% \break
\begin{figure*}[!htbp]
\centerline{\includegraphics[width=\linewidth]{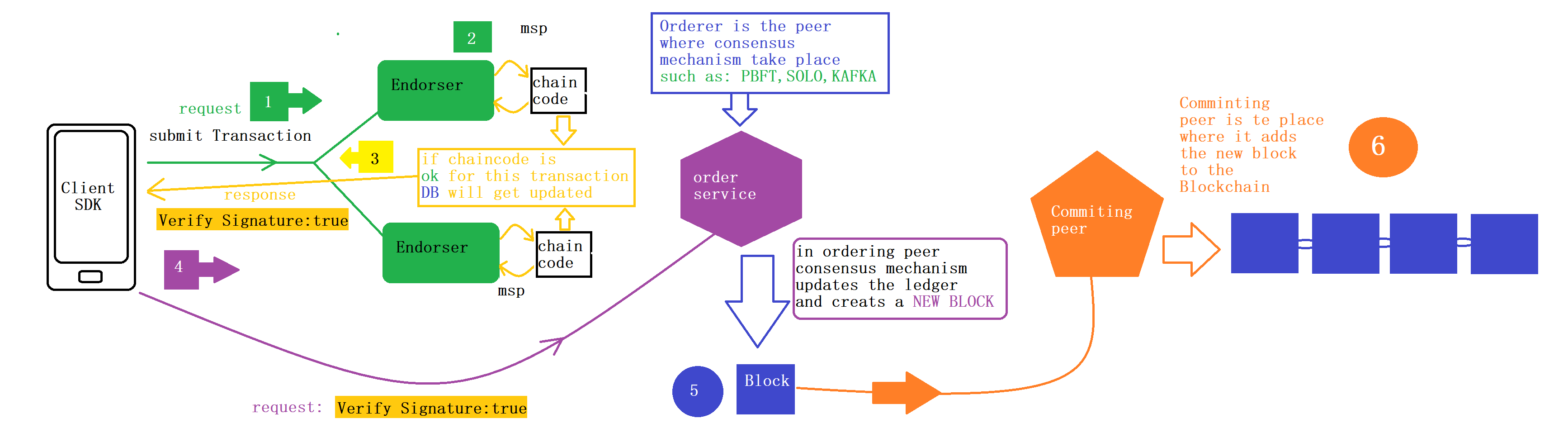}}
\caption{Step by step process of Data flow and Governance with three different peer types inside the Blockchain Network}
\label{fig:favricGovernance}
\end{figure*}

%%%Fig 6
% In the Figure: \ref{fig:favricGovernance} the whole data flow and governance are visually represented. Now we can 
% explain it's every process step by step. In the diagram (Figure: \ref{fig:favricGovernance}) we are showing the 
% 'Endorser' and 'Committing peer' differently to make the idea more clear. But in our 'Proof of Concept' our 'Endorser' and 'Committer' will be the same peer for each organization.
In the Fig: \ref{fig:favricGovernance}, the entire data flow and governance are visually represented.  This Fig exhibits the 'Endorser' and 'Committing Peer' differently to make the idea clearer. But in our implementation, the 'Endorser' and 'Committer' will be the same peer for each organization.

% \begin{enumerate}
%   \item The first step in the Figure: \ref{fig:favricGovernance} is the client application sending a request to the dedicated organization's anchor peer. Which will act as Endorser.
%   \item In this step the endorser will first start the MSP operation to check the user validation. If request comes from a valid user the endorser will execute the chaincode and World State will get updated.
%   \item Then the endorser will send a response to the client side application to confirm that the digital signature is true / valid.
%   \item In this step the change / invoke operation which was executed through the chaincode will be sent to the Ordering peer. Ordering peer will run the Consensus mechanism. To run this operation Ordering peer can be configured with any consensus algorithm. Such as PBFT, Raft, Kafka etc. For our Proof of Concept we will be using Raft. That is why we have multiple ordering peers. So that Raft can elect any Orderer as a leader at any situation to get the job done.
%   \item After a successful Consensus mechanism a new updated block will be created.
%   \item When the Ordering peer confirms the new block, all the committing peers add the block to its blockchain.
% \end{enumerate}

\begin{enumerate} 
\item The first step in the Fig: \ref{fig:favricGovernance} is the client application sending a request to the anchor peer of the corresponding organisation. This will act as the endorser. 
\item In this step, the endorser first starts the MSP operation to check user validation. If the request is from a valid user, the endorser executes the chain code and the world state is updated. 
\item Then, the endorser sends a response to the client-side application to confirm that the digital signature is true/valid. 
\item In this step, the change/ invoke operation performed over the chain code is sent to the ordering peer. The ordering peer will execute the consensus mechanism. To perform this operation, the ordering peer can be configured with any consensus algorithm. Such as PBFT~\cite{pbft}, Raft, Proof of Work (POW)~\cite{consensus}, Proof of Stake (POS)~\cite{consensus}, etc. In the proposed system Raft is used as the consensus mechanism. For this reason, there are multiple ordering peers in the proposed system. So that Raft can elect any orderer to be the leader in any situation to complete the task. 
\item After a successful consensus mechanism, a new updated block is created. 
\item When the ordering peer confirms the new block, all binding peers add the block to their blockchain.
\end{enumerate}

\subsection{Governance}
% So from the explanation of the architecture we can see that our proposed solution maintains a well Governance body. For every operation this system's every node has individual governance role. Every dedicated organization's admin can govern the user enrollment process and generating user's health card with the Digital Signature distributed 
% by the CA-Server (MSP). Endorsing peers will validate user's identification (MSP) and his/her requested operation (chaincode). If these two governance pipeline pass successfully the Orderer authorizes the data and generates a new block. After all these there will be a dedicated peer component for committing the block in every peer of the DLT. In the Figure: \ref{sfig:peerstructure} and \ref{sfig:orderedstructure} it can be seen that every peer and orderer holds the block which is always kept up-to-date. From the Figure: \ref{fig:addUser} it is also visible that every organization's each user is added in the Private DLT by strict governance under its own structure. That is how our Decentralize Ledger Technology will work with a strong governance body which will be technically automated and trusted as the whole governance is neutral.

\begin{figure}[!t]
    \centering
    %\vspace*{-.02in}
    \subfloat[Block location in orderer structure]{
        \hspace*{-.1in}
        \includegraphics[scale=0.25]{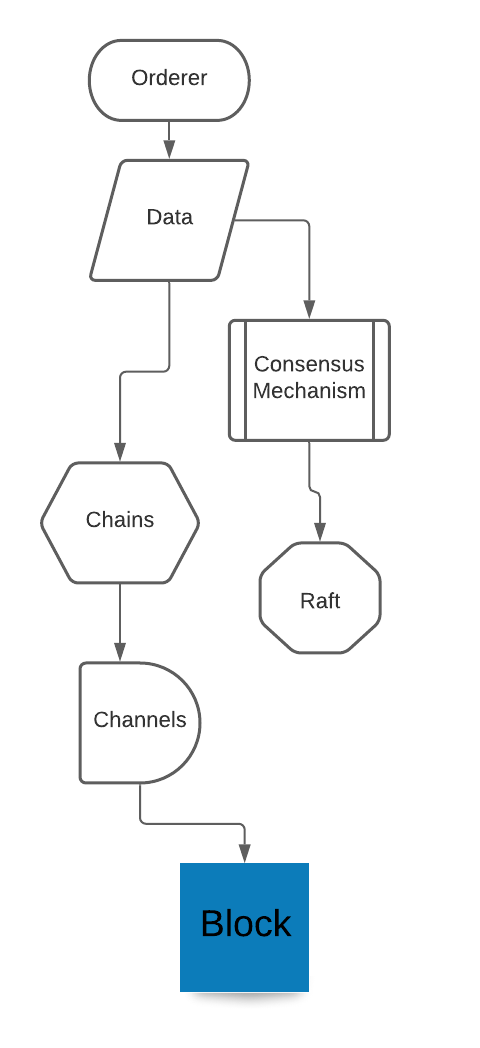}
        \label{sfig:peerstructure}
    }
    %   \vspace*{-.14in}
    \hfill
    \subfloat[Block location in peer structure]{
        \hspace*{-.1in}
        \includegraphics[scale=0.25]{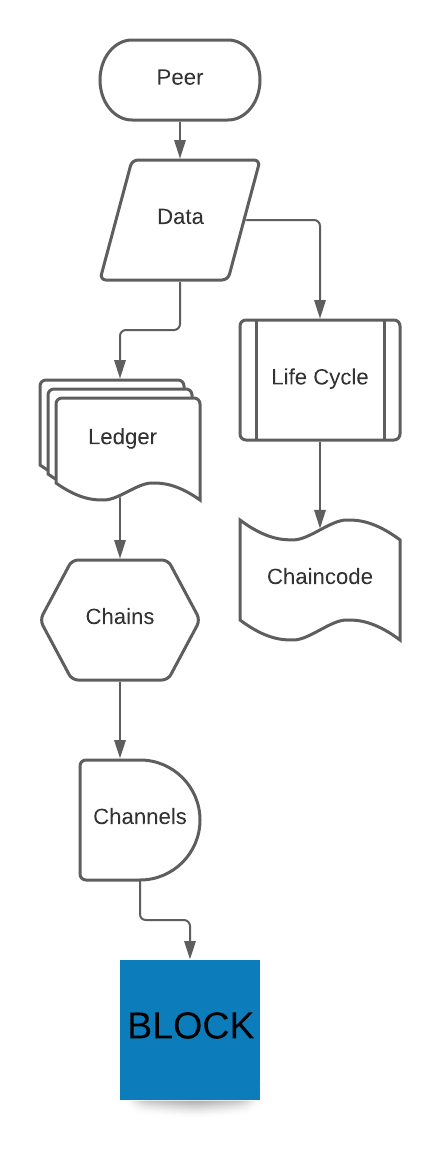}
        \label{sfig:orderedstructure}
    }
    %   \vspace*{-.14in}
    \caption{Location of a Block inside the hyper ledger fabric framework in the orderer and peer structure.}
    
    \label{fig:peerorder}
\end{figure}

From the explanation of the architecture, we can see that the proposed solution maintains a well-governed body. For each process in this system, each node has its own governance role. Any administrator of an organization can control the process of user registration and creation of user's health card with the digital signature distributed by CA server (MSP). The confirming peers validate the user's identification (MSP) and the operation requested by the user (chain code). If these two checks are successful, the orderer authorizes the data and generates a new block. After all this, there is a dedicated peer component in each peer of the DLT for the commitment of the block. In the Fig: \ref{sfig:peerstructure} and \ref{sfig:orderedstructure}, it can be seen that each peer and orderer holds the block which is always up to date. From the Fig: \ref{fig:addUser}, it can also be seen that each user is added to each organization in the private DLT through strict governance under its own structure. In this way, our decentralized ledger technology works with a strong governance body that is technically automated and trustworthy as all the control is neutral.

\subsection{Privacy and Access}
% The blockchain technology based healthcare services may clear obstacles to patients acquiring copies of their healthcare records or transferring them to another healthcare service provider. Records can be verified after being signed by the source and added to the blockchain. This service can  guarantee unalterable patient records with the help of asymmetric key algorithm's key pair concept. Encrypted data in the blockchain can only be read with the patient’s private key, which would empower patients to control access to their sensitive data, which is Consistent with the European General Data Protection Regulation (GDPR), Bangladesh  Digital  Security  Act and other healthcare security regulation (HIPAA) \cite{badr2019blockchain}.

Blockchain-based healthcare services could remove barriers to patients acquiring copies of their health records or transferring them to another healthcare service provider. Records can be verified after they are signed by the source and added to the blockchain. This service can guarantee the immutability of patient records using the key pair concept of the asymmetric key algorithm. Encrypted data in the blockchain can only be read using the patient's private key, which allows patients to control access to their sensitive data. This is in line with the European General Data Protection Regulation (GDPR)~\cite{voigt2017eu}, Bangladesh Digital Security Act \cite{6} and other healthcare security regulations (HIPAA) \cite{badr2019blockchain}.
% \newline

\section{Implementation}
This section describes the implementation of the proposed system.
% In this segment the whole implementing progess will be described according to the System Design section.

\begin{figure*}[!htbp]
\centering
    \includegraphics[scale =.5]{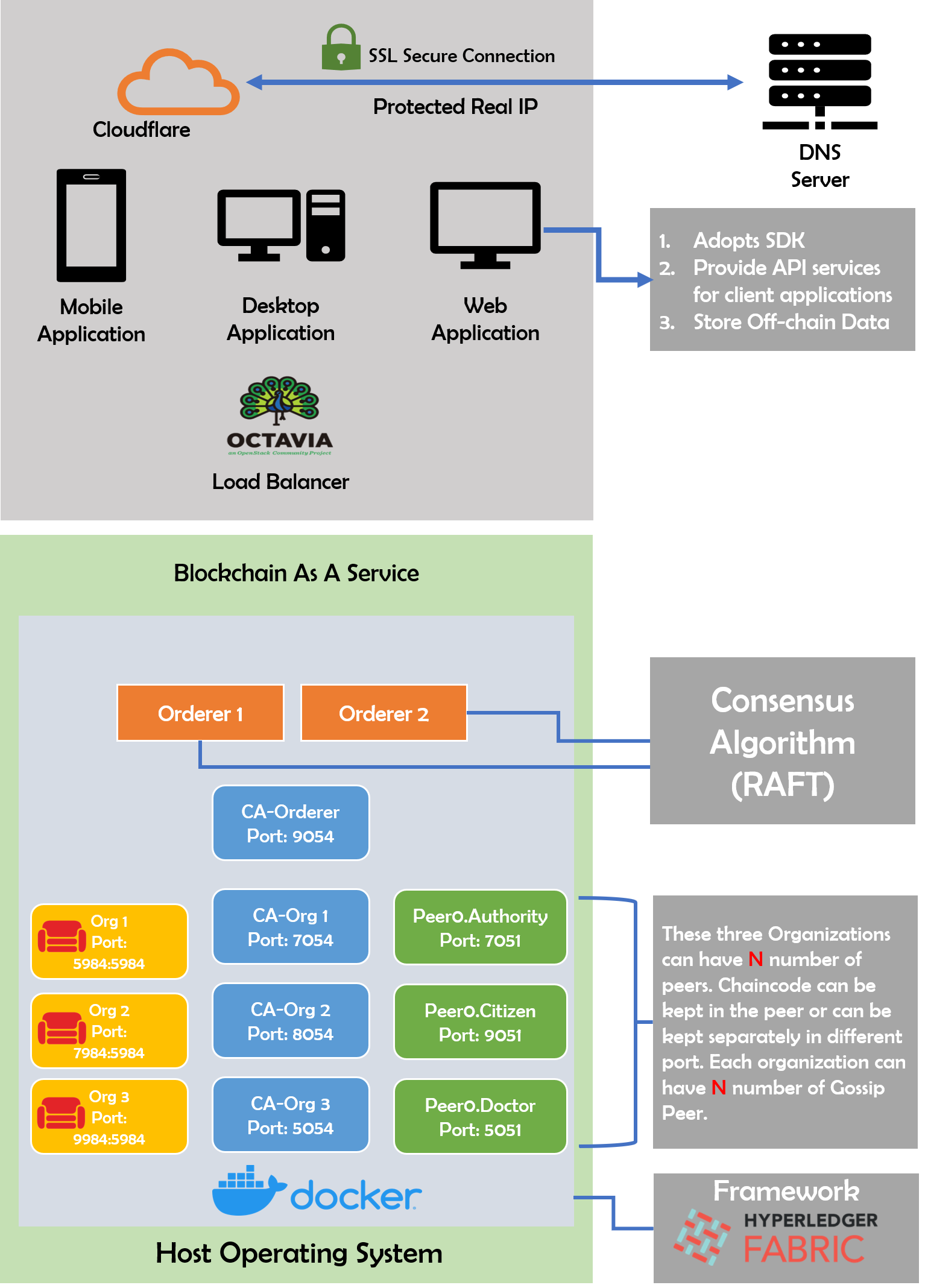}
    \caption{System Architecture (DLT represented with Docker Containers)}
    \label{fig:SystemBody}
\end{figure*}

\subsection{Infrastructure}

% For developing the physical infrastructure our setup is very simple. A fiber optical internet connection provided by the ISP, one Optical Network Unit (ONU), one Wi-Fi Router (2.4 GHz) and one general Desktop Computer for Hosting Server. 
To develop the physical infrastructure, our setup requires some simple components. To set up the proposed system, a stable Internet connection, an optical network unit (ONU), a Wi-Fi router (2.4 GHz), and a generic desktop computer for the hosting server is required. The minimum configuration required for the hosting server is as follows.
\begin{itemize}
  \item 16 GB RAM
  \item 256 GB SSD
  \item Intel Core i7 Processor (7th generation, 4 Cores processor with 8 threads, 3.90GHz Frequency with 8MB Cache)
  \item 650W Power Supply
\end{itemize}

% As this prototype has limited users and traffic there are no extra cooling system for our physical devices. Though this project was live hosted once. So for that our design do have some safety precaution for our hardware safety which will be discussed in the Security segment from Figure: \ref{fig:SystemBody}.  
Since this prototype has a limited number of users and traffic, no additional cooling system is required for the physical devices. However, considering the production level deployment, the proposed system includes some safeguards for the security of the hardware, which are explained in section \ref{ss:security} titled  "Security".

\subsection{Security}
\label{ss:security}
% For hosting this service, a Real IP is assigned to our router. And router was configured for port forwarding. In the configuration our External Port was 80 for Hosting the entire service. To hide the Real IP and port number from the internet world \textit{Cloudflare} is used as CDN service. This CDN service keep protected our IP and redirect a domain name to our hosting server. 
% \newline
% \textit{Cloudflare} also fulfill our some of the Security requirement. Such as \textit{SSL} security protocol. This creates an encrypted link between our web server and client browser. At Figure: \ref{fig:SystemBody} top layer the use of \textit{Cloudflare} is visible. This CDN service also have feature for monitoring and protecting our server from Denial-of-service attack (DDoS). While monitoring the traffic if any abnormal traffic is identified "Under Attack Mode" can be activated for protecting the web server from DDoS attack. \textit{When "Under Attack Mode" is activated the downfall is all API services will become unavailable and web client have to go through a CAPTCHA test while visiting the website.}  
A public IP is assigned to our router for hosting this service. And the router was configured for port forwarding. In the configuration, external port is 80 for hosting the whole service. To hide the public IP and port number from the internet, Cloudflare is used as a CDN service~\cite{dewi2019implementation,siduzzaman2020performance}. This CDN service protects the public IP and redirects a domain name to the hosting server. Cloudflare also fulfills some of the security requirements. For example, the SSL security protocol~\cite{kant2000architectural,rahman2014shimpg}. This establishes an encrypted connection between the web server and the client browser. In Fig \ref{fig:SystemBody}, the use of Cloudflare is visible at the top level. This CDN service also has a feature to monitor and to protect the server from denial of service (DDoS) attacks~\cite{lau2000distributed}. If abnormal traffic is detected during traffic monitoring, "Under Attack Mode" is enabled to protect the web server from DDoS attacks. When "Under Attack Mode" is enabled, all API services become unavailable and Web clients must pass a CAPTCHA~\cite{von2003captcha} test to visit the website.

% Start reading from here. [saef]
\subsection{Network Setup}
% To established the personal server, there are two things needed to be configured. One is a \textit{Router network configuration}
% and another is a \textit{virtual box network configuration}. Several virtual machines must be able to connect to physical and
% virtual networks with their \textit{virtual network adapters} in virtual box GUI. In this process virtual box connected the
% network through different processes. Virtual box provides a lot of different modes of networking and those are:
% \begin{itemize}
% \item Not Attached
% \item NAT
% \item NAT Network
% \item Bridged Adapter
% \item Internal Network
% \item Host-Only Adapter
% \item Generic Driver
% \end{itemize}
% Here, Bridge Adapter and Nat Network is being used for configure the network for application access.
To set up the personal server, two things need to be configured. One is a \textit{router network configuration} and the other is a \textit{virtual box network configuration}. Multiple virtual machines need to be able to connect to physical and virtual networks through their \textit{virtual network adapters} in the Virtual Box GUI. In this process, Virtual Box connects the network through various processes. Virtual Box provides many different types of networking modes, as follows:
\begin{itemize}
\item NAT Network~\cite{bhowmik2020iot}
\item Bridged Adapter~\cite{ali2011virtual,miah2020location}
\item Internal Network
\item Host-Only Adapter
\item Generic driver
\end{itemize}
Here Bridge Adapter is used to configure the network for application access.
% \newline
\subsubsection{Bridge Adapter}
% This mode is used for connecting the virtual network adapter of a VM to a physical network to which a physicalnetwork adapter of the VirtualBox host machine is connected. Through the windows machine (host machine) is connected with the WIFI network. So, those processes are given below that described the bridge network is enabled.
This mode is used to connect the virtual network adapter of a VM to a physical network to which a physical network adapter of the VirtualBox host machine is connected. The  host machine is used to connect to the WIFI network. The following steps describe the process to enable the bridge network.
% \newline
\begin{enumerate}
    \item First need to select the settings option from the selected VM.
    \item Then select the network option from the GUI.
    \item After that, attached to the adapter option of the bridge adapter network mode.
    \item Then the wireless network adapter option needs to be selected through bridge adapter option.
\end{enumerate}

% As, a result the guest machine got the IP which is similar as the host machine(windows). This IP is assigned through the router configuration.
As a result, the guest computer has an IP address that is the same as the host computer. This IP address is assigned via the router configuration.
% \newline
% The router is connected with the real IP or public IP like 103.96.37.122. Now Router assigned several machinesthrough the private IP. Here, 192.168.0.109 IP with port 80 is assigned to the guest machine (Linux) throughout thenetwork mode bridge adapter. But still, no one can access the application with public IP. NAT (Network Address Translation) is not configured in the router to reach the virtual machine.
The router is connected to  the public IP like 103.96.37.122. Now the router has assigned multiple private IP addresses to  multiple machines . Here the IP 192.168.0.109 with port 80 is assigned to the guest machine (Linux) through the bridge adapter in network mode. However, no one can access the application with the public IP. NAT (Network Address Translation) is not configured in the router to reach the virtual machine.
\newline
% So, configure the router need to access the interface web page of the router. Those are accessed by the IP 192.168.0.1. After that, passing the login page the GUI of router configuration has appeared. Then clicked the Nat forwarding option. After that, click the Virtual server and provide the external port and internal port with the private IP of the guest machine. As a result, the port forwarding is completed. Now the personal server setup is completed and accesses the application globally by the public IP.

To configure the router, it  needs to access the router's interface. This interface is accessed via the IP 192.168.0.1. After passing the login page, the GUI of the router configuration appears infront of the user. Then, select  the NAT forwarding option. Next, select Virtual Server and specify the external port and the internal port with the private IP of the guest machine. This completes the port forwarding mechanism. Now the personal server setup is complete and the application is  open to be accessed globally using the public IP.

\subsection{Application layer}
% In the top layer of Figure \ref{fig:SystemBody} three different platform are mentioned which are Mobile, Computer and Web applications. This system have this cross platform ability by adopting SDK and API Gateway.
% \newline
% As mentioned before we are following Hyperledger Fabric Framework and will be using JavaScript SDK we have maintained some  Object-oriented programming (OOP) standards. \textit{Inheritance} is widely used here for adopting SDK components in our NodeJS environment.
% \newline

% \textbf{fabric-contract-api}, \textbf{fabric-ca-client}  and \textbf{fabric-network} these three modules from Hyperledger Fabric are most commonly used during the implementation of our project. 
In the top level of Fig \ref{fig:SystemBody}, three different platforms are mentioned, namely mobile, desktop and web applications. This system supports  cross-platform capability by using SDK and API gateway.
As mentioned earlier, the proposed system uses Hyperledger Fabric Framework and JavaScript SDK while maintaining Object Oriented Programming (OOP) standards. \textit{inheritance} is largely used here for adopting SDK components into our NodeJS environment.
\textbf{"fabric-contract-api"}~\cite{fcontract}, \textbf{"fabric-ca-client"}~\cite{fca} and \textbf{"fabric-network"}~\cite{fnetwork} these three modules of Hyperledger Fabric are mostly used in the implementation of the prototype system.

\subsubsection{fabric-contract-api}
This module is used when a smart contract needs to be installed in a peer or for initializing the smart contract / chain code.  

\subsubsection{fabric-network}
% This is one of the most important module which connects the Digital \textit{Healthcard} (knows as Wallet in Fabric) with our Application Layer. The Figure \ref{fig:userNdlt} is the complete output of this module. From enrolling an user in the private-blockchain network to creating a \textit{Healthcard} for that user, this module is widely used. This is also responsible for verifying the \textit{Healthcard}. Figure \ref{fig:addUser} represents the complete work flow of this user enrollment operation.
% \newline
% This module also have another object named \textit{gateway}. This object connects an user to the blockchain network which returns a \textit{contract} (object). It helps the client / user to run his / her read-write operations in the smart contract.
% \newline
% Even the user login authentication process are operated from here. At Figure \ref{fig:LoginSDK} this process is shown step by step for better understanding. 
This is one of the most important modules that connects the digital \textit{healthcard} (known as a wallet in Fabric) to the application layer of the proposed system. The Fig \ref{fig:userNdlt} represents the overview of this module . From logging a user into the private blockchain network to creating a \textit{healthcard} for that user, this module is widely used. It is also responsible for verifying the \textit{healthcard}. Fig \ref{fig:addUser} represents the complete workflow of this user registration operation.
This module also has another object called \textit{gateway}. This object connects a user to the blockchain network which returns a \textit{contract} object. It helps the client / user to perform its read and write operations in the smart contract. The authentication process for user login is also controlled from here. Fig \ref{fig:LoginSDK} exhibits this process step by step for better understanding.

\subsubsection{fabric-ca-client} 
% This module is responsible for establishing communication with the CA-Server. Whenever \textbf{fabric-network} is used through the \textit{Healthcard} \textbf{fabric-ca-client} module is called every time for authenticating the used and its request. 

% All these process are managing from the Node.js environment which is running at internal port 3000. In the Node.js environment the main web application which is actually dealing with the SDK is developed with \textbf{Express} framework which is actually a \textit{MVC} based web application framework. This web application is actually acting as a \textit{Portal} between client applications and blockchain service (Figure \ref{fig:userNdlt}). This Web application has its own client interface and also API service. \textit{This is the actual Web service which is redirected from the Router at External port 80}.
% \newline
% For patient's flexibility a mobile application version is also developed for our blockchain service. For that \textbf{Flutter} is used. This \textbf{Flutter} based mobile application (Android/IOS) is using those API services which are provided by the \textbf{Express} web application.
% \newline
% Doctor's official Software is developed with \textbf{C\#} which is actually a \textit{Windows} based Computer Software using the same API services provided by the \textbf{Express} based web application for accessing the blockchain services.  
This module is responsible for establishing communication with the CA server. Whenever \textbf{fabric-network} is used over the \textit{healthcard}, the module \textbf{fabric-ca-client} is called each time to authenticate the user and his request. 

All these processes are managed by the Node.js environment running on internal port 3000. In the Node.js environment, the main web application that actually works with the SDK is developed using the \textbf{Express} framework, which is actually an MVC-based web application framework. This web application acts as a portal between client applications and blockchain service. This web application has its own client interface and also an API service. This is the actual web service that is redirected by the router to the external port 80.
For patient flexibility, a mobile application for our blockchain service is also developed for Android and iOS platforms. For this purpose, flutter~\cite{flutter} is used. This mobile application uses the API services provided by the Express web application~\cite{brown2019web}.
The physicians' software is developed using C\#, a Windows-based desktop application software that uses the same API services provided by the Express-based web application to access the blockchain services.

\subsection{blockchain as a service}
% This section discusses about the architecture and implementation of the "blockchain As A Service" portion of the proposed system. Figure \ref{fig:SystemBody} represents the overview of the proposed system.
This section describes the architecture and implementation of the blockchain As A Service part of the proposed system. Fig \ref{fig:SystemBody} presents the overview of the proposed system.

\subsubsection{Framework}
% For robust privacy and security features, with support for granular access control, private channels and well documentation we have chosen Hyperledger Fabric for implementing this project \cite{cbfha2019Wiley}. The framework version v2.1.0 is used for development.

% This framework have same pre-built \textit{docker images} which will help the development process to ensure all the blockchain standards. Image
% is a ready-made read-only template that creates the container. At Figure \ref{fig:SystemBody} it can be seen that the Fabric framework actually holds the whole blockchain service with many docker containers. 

For robust privacy and security features with support for granular access control, private channels, and good documentation, we chose Hyperledger Fabric to implement this system \cite{cbfha2019Wiley}. The framework version v2.1.0 is utilized for the system development.

This framework has pre-built \textit{docker images}~\cite{docker} to support the development process and ensure all blockchain standards. Pre-built docker images are read-only template that creates docker containers with different configuration already implemented on those containers . Fig \ref{fig:SystemBody} shows the architecture of the Fabric framework, which contains the entire blockchain service with many Docker containers.

%%%%%%%%%%%%%%%%%%%%%%%%%%%%%%%%%%%%%%%%%%%%%%%%%%%%%%%%%%%%%%%%%%%%%%%%%%%%%%%%% 
\subsubsection{Virtual Machine Environment}
% To run the blockchain application it's easier to consider a Linux-based operating system from the developer's
% priority. Also, the hyperledger framework is configured based upon Linux OS based. So, the first task was to
% make an environment based upon Linux based. As a Windows-based user, it's needed a virtual machine for
% making the environment setup. Here, \textit{Oracle Virtual box} is used for running the Linux operating system. Then
% all the necessary files like docker, hyperledger fabric framework are set up to the OS which is inside the
% virtual box.
% \newline
% Though all these orderer have their own domain name (Figure \ref{fig:generalArchitecture}). But as this implementation is at prototype level all these individual domain names are identical only at the docker environment. All these blockchain peers are decentralized virtually at one single physical machine and the full blockchain service is live hosted at one single domain name through the router's external port80.  
% \newline
% In the Figure \ref{fig:SystemBody} the whole "blockchain As A Service" is inside a VM including the \textit{Host OS}.

To run the blockchain application, it is easier to use a Linux-based operating system. The Hyperledger framework is also configured based on Linux OS. So the first task is to create a Linux-based environment.  A virtual machine environment is required to run a Linux-based guest operating system from windows based host operating system. In the proposed system, \textit{Oracle Virtual box}~\cite{vb} virtual machine platform is used to set up the Linux operating system on the Windows-based machine. Then, all the required applications such as Docker and Hyperledger Fabric Framework are set up on the Linux-based OS residing in the virtual machine environment.

It is true that all these clients have their own domain name. However, since this implementation is a prototype, the individual domain names are identical only in the Docker environment. All of these blockchain peers are virtually decentralized on a single physical machine and the entire blockchain service is hosted live under a single domain name, using the router's external port 80. 

Fig \ref{fig:SystemBody} shows that the entire "blockchain As A Service" part is implemented in a VM environment.

\subsubsection{Docker Container}
% Inside the VM Docker is used to automate the deployment of the application as a lightweight container so that the application can work efficiently in different environments. It provides a lot of processes with virtualization system architecture. As a result, it's easy to deploy an application in a decentralized architecture in prototype level blockchain service.
Within the VM, Docker is used to automate the deployment of the application as a lightweight container, allowing the application to work efficiently in different environments. It provides a variety of processes with virtualization system architecture. This makes it easy to deploy an application in a decentralized architecture in a prototype-level blockchain service.

\subsubsection{CA Server}
% In this permission based Private blockchain network \textit{CA Server} or Certificate Authority Server is the most important part. This dedicated service is responsible to identify and certify each orderer, organizations and also its each user. For implementing this project this services are also hosted at our one single physical machine's different docker containers. At Figure \ref{fig:SystemBody} we can see that there are 4 (four) containers hosting these servers in four different ports (9054. 7054, 8054, 5054). In the Figure \ref{fig:SystemBody} it can be seen that there is one dedicated CA Server (port: 9054) for certifying the orderder peers and the other three organizations have also their own CA Server.
% Each peer admin and user must get certified and verified through the dedicated CA Server where they belong.

In this permission-based private blockchain network, \textit{ CA server or Certificate Authority Server} is the most important part. This particular service is responsible for identifying and certifying each orderer, organization and also each user. For the implementation of this system, these services are also hosted in different Docker containers on a single physical machine. In Fig \ref{fig:SystemBody}, we can see that there are four containers hosting these servers in four different ports (9054, 7054, 8054, 5054). It can be seen that there is a dedicated CA server (port: 9054) for certifying the orderder peers and the other three organizations also have their own CA server. Each peer admin and user has to get certified and verified through the dedicated CA server they belong to.

\subsubsection{World State}
% In the Figure \ref{fig:SystemBody} docker's yellow labeled containers are World State of our system. These World State are also running on different docker containers which are operating at different ports (5984, 7984, 9984) identified as Org1, Org2 and Org3. Only the Organization peer have their own World State for fast query and client support. 
In the Fig \ref{fig:SystemBody}, the Docker containers highlighted in yellow are the World State of the proposed system. These World states also run on different Docker containers running on different ports (5984, 7984, 9984) identified as Org1, Org2 and Org3. Only the organization peers have their own World State for fast queries and client support.
\subsubsection{Orderer}
% In case of Orderer in our Private blockchain network 'N' number of Orderer can be created. Because after each transaction in any of the peer orderer is the actual Private blockchain authority which automates the verification process in the system. 
% \newline
% In this project implementation we have created only 2 (two) orderers in our system. At Figure \ref{fig:SystemBody} those are labeled with Orange color (Orderer 1 and Orderer 2).These two orderer are running at Port 7050 and 8050 (Figure \ref{fig:generalArchitecture}). 
In the case of Orderer in our private blockchain network, an 'N' number of Orderer can be created. Because after each transaction in one of the peer orderers is the actual Private blockchain authority that automates the verification process in the system. 
In this system implementation, we have created only 2 (two) orderers inside our system. In Fig \ref{fig:SystemBody}, these are marked with orange color (Orderer 1 and Orderer 2). These two orderers run on port 7050 and 8050. 

%\subsubsection{Organization Peer}
\subsubsection{User Enrollment Process}
% The implementation process at new Client enrollment is a very interesting part. Because at Private blockchain network preventing the bad actor and keep every actor / visitors identical is a very important thing.
% \newline
% So for ensuring these standards Figure \ref{fig:addUser} has been followed during implementation. Here \textit{CA Server} is performing as "Membership Service Provider" (MSP). Before that it should be mentioned that every Organization which are part of this blockchain service has a \textbf{Default Admin}. This role known as \textbf{Admin} is responsible for authorizing a new user in its dedicated organization. And if a user is added he / she must have a dedicated \textit{Digital Health Card} for the further existence in the system (Figure \ref{fig:userNdlt}).
% \newline
% \textit{Note that an \textbf{Admin} can enroll an \textbf{User} and \textbf{Admin} both in the system. But the \textbf{User} can only perform its general operations in the system.}
% \newline
% At Figure \ref{fig:addUser} it can be seen that Organization \textbf{X} has an \textbf{Admin} which is enrolling a new \textbf{User} where \textit{CA Server} is serving as \textit{MSP}. Through this collaboration a new \textit{Health Card} has been generated for a new user.
% \newline
% \textit{This is how an actor can not get access or leave from a Private blockchain network by his / her own will}.
The implementation process in registering new clients is a very interesting part. Because with a private blockchain network, it is very important to prevent bad actors and keep all actors/visitors identical.
To ensure these standards, processes depicted in Fig \ref{fig:addUser} is followed in the implementation. Here, \textit{ CA server} acts as a "Membership Service Provider" (MSP). Each organization that is part of this blockchain service has a \textbf{Default Admin}. This role, known as \textbf{Admin}, is responsible for authorizing a new user in the corresponding organization. And when a user is added, he/she must have a special \textit{Digital Health Card} for continued existence in the system.
 An \textbf{admin} can register both an \textbf{user} and an \textbf{admin} user in the system. But an \textbf{user} can only perform its general operations in the system.
From Fig \ref{fig:addUser}, it can be seen that the \textbf{X} organization has an \textbf{Admin} that signs up a new \textbf{User}, with \textit{ CA Server} serving as \textit{MSP}. This collaboration has created a new \textit{healthcard} for a new user. This mechanism implies that  an actor cannot gain access to or leave a private blockchain network of its own volition.

% Admin & Use enrollment process is same
%Users will be addded by Admin
\subsubsection{Login Authentication Process}
%  User Login authentication is that stage where the whole implementation process took a new turn. Though \textit{Hyperledger Fabric} have a very good documentation support. But in case of merging a MVC framework (Client Portal at Figure \ref{fig:userNdlt}) with a whole new technology (blockchain) for a better human integration experience there were so much difficulties during development. For the lack of proper documentation this segment was self engineered by us in most cases.  
%  \newline
%  The whole process of login and authentication are described at Figure \ref{fig:LoginSDK} with a flowchart.
%  \begin{enumerate}
%     \item Start
%     \item Get credentials(Identity and password) from user as inputs
%     \item check if identity exists in health-card \\ if NO then print "Invalid Identity" \\ else \\ get password from World-state using identity
%     \item check if password matches \\ if YES then give access \\ else \\ print "Invalid Password"
%     \item End
%  \end{enumerate}
%  This whole operation are managed with the help of SDK where \textit{Health Card} are the key to access the blockchain network. 
%  \newline

\begin{figure}[!htbp]
    \includegraphics[width =\linewidth]{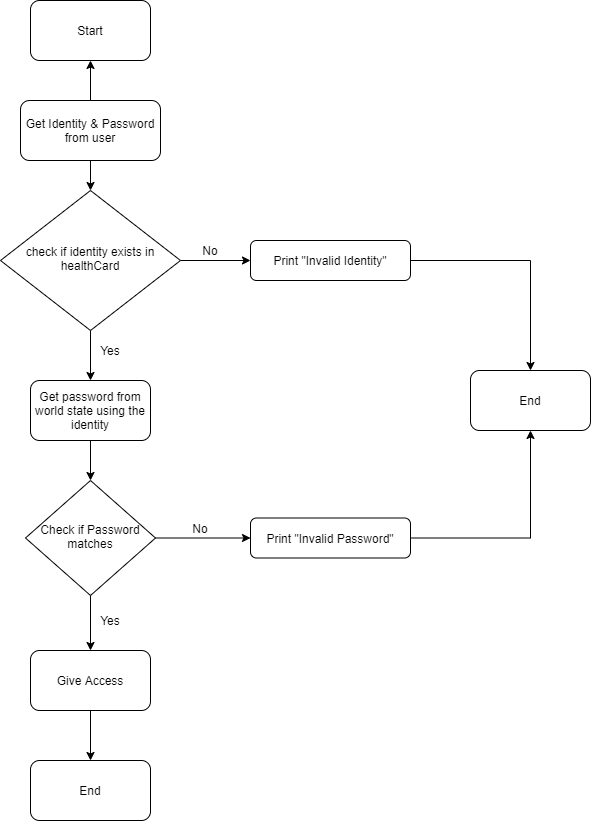}
    \caption{User Authentication Process Diagram (SDK)}
    \label{fig:LoginSDK}
\end{figure}

User login authentication is the phase where the whole implementation process took a new turn. although the \textit{ Hyperledger Fabric} has very good documentation. But in merging an MVC framework with a completely new technology (blockchain) for a better integration experience, there were so many difficulties faced in the development process. In the absence of proper documentation, this area was mostly developed by ourselves. 
The whole process of login and authentication is described in Fig \ref{fig:LoginSDK} with a flowchart.
 \begin{enumerate} \item Start \item Get credentials (identity and password) from user as inputs. \item check if identity exists in health card \ if NO then print "Invalid Identity" \\\ else \\ get password from World-state using identity. \item check if password matches \ if YES then grant access \ otherwise \ print "Invalid Password" \item end
 \end{enumerate}
 This whole operation is managed using the SDK, where \textit{Health Card} is the key to access the blockchain network.
 %%%%%%%%%%%%%%%%%%%%%%%%%%%%%%%%%%%%%%%%%%%%%%%%%%%%%%%%%%%%%%%%%%%%%%%%%%%%%%%%%%%%%%%%%%%%%%%%%%%%%%%%%%%%%%%%%%%%%%%%%%%%%%%%%%%%%%%%%%%%%%%%%%%%%%%%%%%%%%%%%%%%%%%%%%%%%%%%%%%%%%%%%%%%%%%%%%%%%%%%%%%%%%%%%%%%%%%%%%%%%%%%%%%%%%%%%%%%%%%%%%%%%%%%%%%%%%%%%%%%%%%%%%%%%%%%%%%%%%%
 
 %Sir Please check this statement. Suggest me a suitable place for its placement.
%  \textcolor{red}{The downfall of this arrangement is that some of the advantages of blockchain are lost; for example, data redundancy which helps to protect against security attacks o n data availability (eg,Ransomware and denial of service attacks) are lost with off-chain storage.\cite{cbfha2019Wiley}}

%%%--------------------------------------------------------

\section{Discussion}
% The National ID Card (Bangladesh) is a perfect example to proof the point. As we have seen that the whole nation already have a unique identity. At least those who don't have their NID can be uniquely identified with their Birth Certificate. And for operational example all the Banks of Bangladesh already operating with their customer with any of these two documents. The government is sharing their web services to verify these documents through online for these financial organizations. In some cases APIs are also shared by the government for this services. So we can claim that one of the big challenges of adoption permissioned blockchain has been solved already.
% \newline
% The positive side is that we don't need any invest to enroll every citizen in our system again. Simple collaboration from “National Identity Wing” can save investment and time both. This is the way by which we will be able to bring not only the health officials but also the whole nation under our health network.
The National ID Card of Bangladesh is a perfect example to prove that we are ready to adopt private blockchain application in national level. As we have seen, the whole nation already has a unique identity. At least those who do not have their NID can be uniquely identified by their birth certificate. And as a practical example, all banks in Bangladesh are already working with their customers with one of these two documents. The government provides its web services to verify these documents online for the financial institutions. In some cases, APIs are also provided by the government for these services. So, we can say that one of the major challenges in implementing permissioned blockchain has already been solved.
On the bright side, we no longer need to invest to register every citizen in our system. Simply working with the "National Identity Wing" can save both investment and time. In this way, we will be able to include not only health officials, but the whole nation in our health network.

% For now in Bangladesh there is no competitor against our proposed solution. But as the 
% demand of DLT is increasing day by day in future this ministry might need this type of technology to adopt in their office work. So if the ministry judge our proposed solution as a under developed product and wait for any well-known third party company to come up with an enterprise standard service that can be a competition for us. But that 
% can be a risk and violation of our National Data Security Policy. Because at that point all of the government healthcare official's information and citizen's medical data will be on the third party's hand. But our proposed solution and PoC is not designed for Competition and Business. It is for the national demand to advancing the technology for Humanity.
% By applying more R\&D on this PoC the Government of Bangladesh can form its own DLT based Healthcare service which can be established and maintained in the national fourth-tier data center which can be a step towards the Cloud-based National blockchain Platform program \cite{6} of Bangladesh. In this paper's “Why our Solution” segment this point is broadly explained.
At the moment, there is no competitor to our proposed solution in Bangladesh. But as the demand for DLT is increasing day by day, the ministry may need this kind of technology for its office work in future. So if the ministry considers the proposed solution as an underdeveloped product and waits for a well-known third party company to provide a standard enterprise service, it could be a competition for us. But that could be a risk, a breach of our national data security policy. Because then all the information of government health officials and citizens' medical records will end up in the hands of third-party vendors. But our proposed solution and PoC is not about competition and Business. It is about the national demand to advance the technology for humanity.
By applying more R\&D to this PoC, the government of Bangladesh can build its own DLT-based healthcare service that can be set up and maintained in the fourth-tier national data center, which can be a step towards the cloud-based national blockchain platform program \cite{6} of Bangladesh. In the "Why our solution" section of this paper, this point is explained in detail.

%%%--------------------------------------------------------

The proposed blockchain application is tested using a load testing approach. The testing and analysis followed the APDEX (Application Performance Index), an open standard developed by an alliance of companies to measure the performance of software applications in the computing domain. Its purpose is to transform measurements into user satisfaction insights. This is achieved by establishing a consistent method for analyzing and reporting on the extent to which measured performance meets user expectations, based on the number of satisfied, tolerant, and frustrated users. The blockchain application is hosted on a server with limited hardware resources. We used a virtual environment with 8 GB of RAM and 2 processor cores. The tests are performed using JMeter tool, the number of concurrent users is set to 100 (threads), the ramp-up time is 10 and the user types are BMDC (superadmin of the system) and Nagorik (citizen). The ramp up time specifies how long JMeter takes to ramp up or activate specific users (threads) for testing. In this scenario, after 10 seconds, all 100 users will be active in our blockchain environment.

\begin{figure}[!htbp]
    \includegraphics[width =\linewidth]{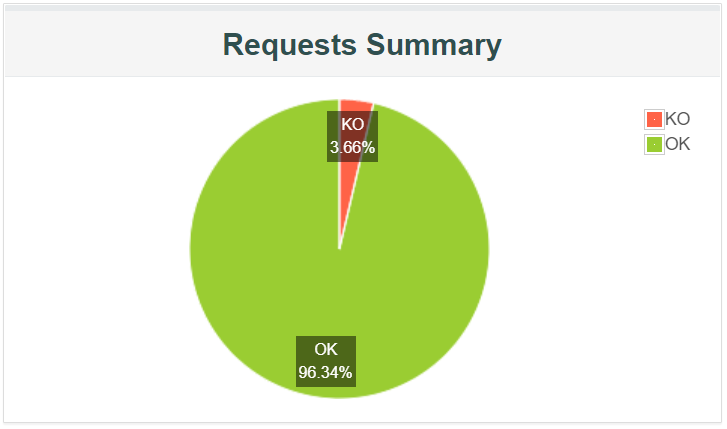}
    \caption{Performance Test Request Summary}
    \label{fig:requestSummary}
\end{figure}

%%%--------------------------------------------------------
The graph shown in Figure \ref{fig:requestSummary} shows the result of summarizing the requests, whether they are in order or not. The graph shows that the results are overwhelmingly positive, with a percentage of 96.34\% positive requests and only 3.66\% negative requests.

\begin{figure}[!htbp]
    \includegraphics[width =\linewidth]{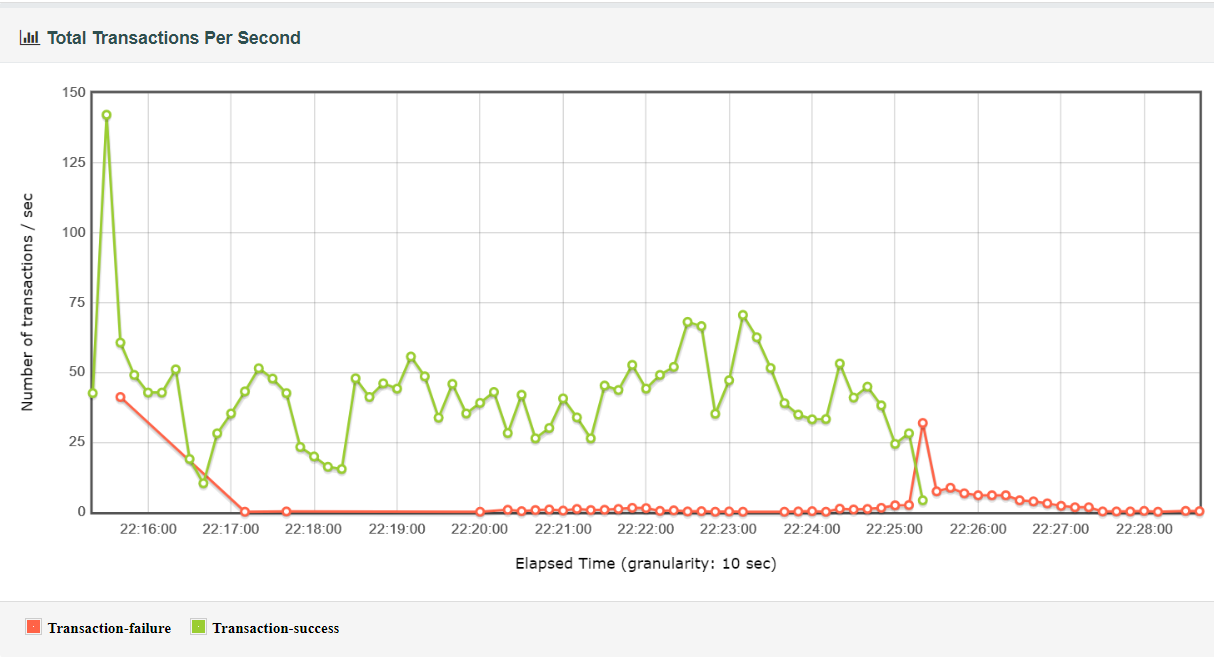}
    \caption{Success and Failure rate of Total Transactions Per Second in granularity of 10 seconds}
    \label{fig:totalTransaction}
\end{figure}

The graph in Figure \ref{fig:totalTransaction} shows the granularity of 10 seconds of total user transactions per second. We can also see that there are only two points in a second where the number of transaction failures is highest. However, the number of failures does not exceed 41 transactions, while the successful transaction rate is 142 per second. So the failure rate of the overall result is very low compared to the success rate.

%%%--------------------------------------------------------

\begin{figure*}[!htbp]
\centerline{\includegraphics[width=\linewidth]{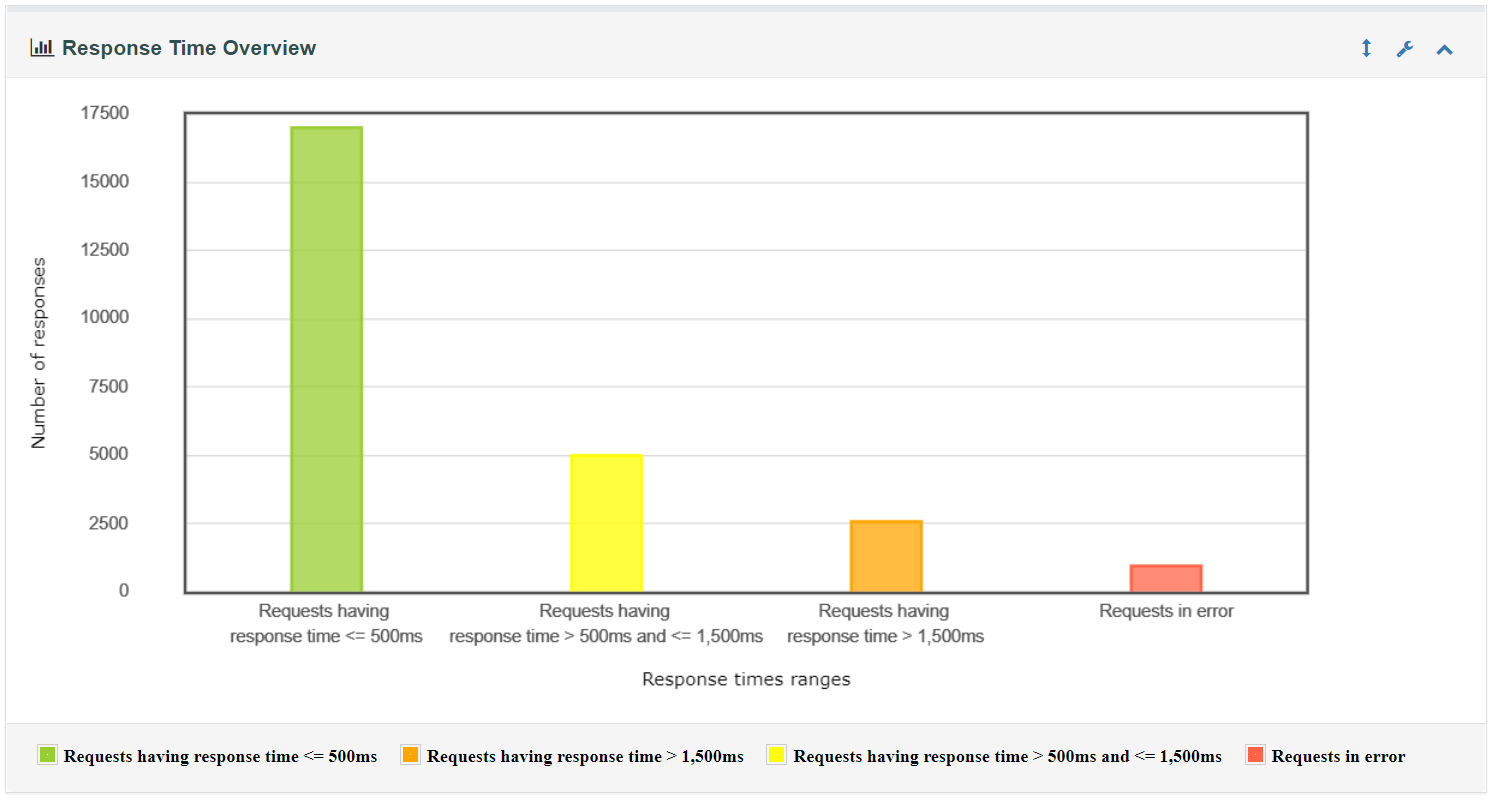}}
\caption{Response Time Overview}
\label{fig:responseTime}
\end{figure*}

The diagram in Figure \ref{fig:responseTime} illustrates the overview of the response times of the requests generated by the users. It can be seen that the majority of the requests have a response time of less than 500 ms, namely 17020 out of 25,509 requests, 4987 requests have a response time of more than 500 ms and less than 1500 ms, 2568 requests have a response time of more than 1500 ms and very few requests failed, namely 934 out of a total of 25,509 requests. Overall, the result looks very positive

It should be noted that these analyses should be considered hypothetical, as our blockchain application is developed and installed on our own configured server. It is therefore subject to hardware and network limitations.

\begin{table}[!htbp]
    \centering
    \scalebox{0.5}{
    \begin{tabular}{|c|c|c|c|c|c|}
         \hline
         \multicolumn{6}{|c|}{Human Resource For converting prototype level to Production level work} \\
         \hline
         \multicolumn{6}{|c|}{Sprint: 20 | Working days in a week: 5} \\
         \hline
         \begin{tabular}{@{}c@{}}Resource \\ Type\end{tabular} & 
         \begin{tabular}{@{}c@{}}Person \\ Count\end{tabular} & 
         \begin{tabular}{@{}c@{}}Experience \\ Level\end{tabular} & 
         \begin{tabular}{@{}c@{}}Salary \\ (BDT/mon)\end{tabular} & 
         \begin{tabular}{@{}c@{}}Contact \\ Duration\end{tabular} & 
         \begin{tabular}{@{}c@{}}Total Cost \\ (BDT)\end{tabular} \\
         \hline
         HR Executive & 1 & Mid & 50,000 & 9 & 450,000 \\ 
         \hline
         Project Manager & 1 & Senior & 150,000 & 9 & 1,350,000 \\ 
         \hline
         UI/UX Designer & 1 & Mid & 50,000 & 3 & 150,000 \\
         \hline
         \multirow{2}{*}{Software Engineer} &
         1 & Senior & 150,000 & 9 & 1,350,000 \\
         & 2 & Mid & 73,000 & 9 & 1,314,000 \\ 
         \hline
         \multirow{2}{*}{Software Developer} &
         2 & Mid & 55,000 & 9 & 990,000 \\
         & 5 & Associate & 35,000 & 9 & 1,575,000 \\ 
         \hline
         \multirow{2}{*}{SQA Engineer} &
         1 & Senior & 70,000 & 9 & 630,000 \\
         & 2 & Associate & 35,000 & 4 & 280,000 \\ 
         \hline
         \begin{tabular}{@{}c@{}}Automation \\ QA Engineer \end{tabular} & 1 & Mid & 55,000 & 4 & 220,000 \\
         \hline
         \multirow{2}{*}{DevOPs Engineer} &
         1 & Senior & 170,000 & 4 & 680,000 \\
         & 1 & Associate & 35,000 & 4 & 140,000 \\ 
         \hline
         Total human resource & \multicolumn{2}{c|}{19} & \multicolumn{2}{c|}{Total Sprint Cost} & 9,129,000 \\
         \hline
    \end{tabular}}
    \caption{Human resource and cost required for the initial development phase.}
    \label{tab:hrdevcost}
\end{table}

\begin{table}[!htbp]
    \centering
    \scalebox{.55}{
    \begin{tabular}{|c|c|c|c|c|}
         \hline
         \multicolumn{5}{|c|}{Human Resource for Maintenance} \\
         \hline
         \multicolumn{5}{|c|}{Working days in a week: 5 | Yearly Bonus: 2} \\
         \hline
         \begin{tabular}{@{}c@{}}Resource \\ Type\end{tabular} & \begin{tabular}{@{}c@{}}Person \\ Count\end{tabular} & \begin{tabular}{@{}c@{}}Experience \\ Level\end{tabular} & \begin{tabular}{@{}c@{}}Salary \\ (BDT/mon)\end{tabular} & \begin{tabular}{@{}c@{}}Total Cost \\ (BDT)\end{tabular} \\
         \hline
         HR Executive & 1 & Mid & 50,000 & 50,000 \\ \hline
         \begin{tabular}{@{}c@{}}Business \\ Administration\end{tabular} & 1 & Senior & 65,000 & 65,000 \\ \hline
         Digital Marketer & 1 & Mid & 35,000 & 35,000 \\ \hline
         Graphics Designer & 1 & Mid & 40,000 & 40,000 \\ \hline
         Project Manager & 1 & Mid & 120,000 & 120,000 \\ \hline
         \multirow{2}{*}{Software Engineer} &
         1 & Senior & 150,000 & 150,000 \\
         & 1 & Mid & 73,000 & 73,000 \\ \hline
         \multirow{2}{*}{Software Developer} &
         1 & Mid & 70,000 & 70,000 \\
         & 2 & Associate & 35,000 & 70,000 \\ \hline
         SQA Engineer & 1 & Mid & 55,000 & 55,000 \\ \hline
         DevOPs Engineer & 1 & Senior & 170,000 & 170,000 \\ \hline
         Total human resource & \multicolumn{2}{c|}{12} & \begin{tabular}{@{}c@{}}Total \\ Monthly Cost\end{tabular} & 898,000 \\
         \hline
    \end{tabular}}
    \caption{Human resource and cost required for the maintenance phase.}
    \label{tab:hrmaincost}
\end{table}

As stated in this article, we have developed a blockchain-based application for our proposed solution that is fully functional. Which ensures all requirements and standards for a Blockchain application. Therefore, in Table \ref{tab:hrdevcost}, blockchain infrastructure is not emphasized separately. Here, we believe it is prudent to invest primarily in client-side applications. This is due to the fact that, despite the fact that the entirety of the research in this article is devoted to blockchain, no detailed work has been conducted on all the necessary features of its user organizations, including the user experience of its users. Therefore, if action is required at the production level, it will be in each of these sectors. Therefore, Blockchain Expert is not allocated a specific budget in our future cost plan.

The Table \ref{tab:hrdevcost} is for the initial approach to the development of the project at the production level. Therefore, there is no permanent position listed in this table. This table depicts an estimate of a 20-sprint project plan in which various human resources will be added at varying intervals. For example, UI/UX designer, Who will be added to the sprint from the very beginning in order to create an accurate visualization of the features and a demo of their user experience. However, he or she will focus on the sprint for the initial three months. In contrast, a DevOps engineer will be added to the sprint for the last four months if we takeș it into account. As they are primarily responsible for server-side work and project deployment. For the duration of the sprint, however, there will be a dedicated management team, developer team, and testing team. As these teams are available for the duration of the sprint, it is simple to select any Software Development Life Cycle (SDLC) Model for the development of this project.

After the proposed solution has been implemented on the production server, a full-time technical team is required which is presented in Table \ref{tab:hrmaincost} with proper cost estimation. Here, we have scaled the technical team's budget to business criteria. Such as Business Administrator, Digital Marketer, and Graphic Designer for the design of content to attract a large audience to this system.

\begin{figure}[!htbp]
    \includegraphics[width =\linewidth]{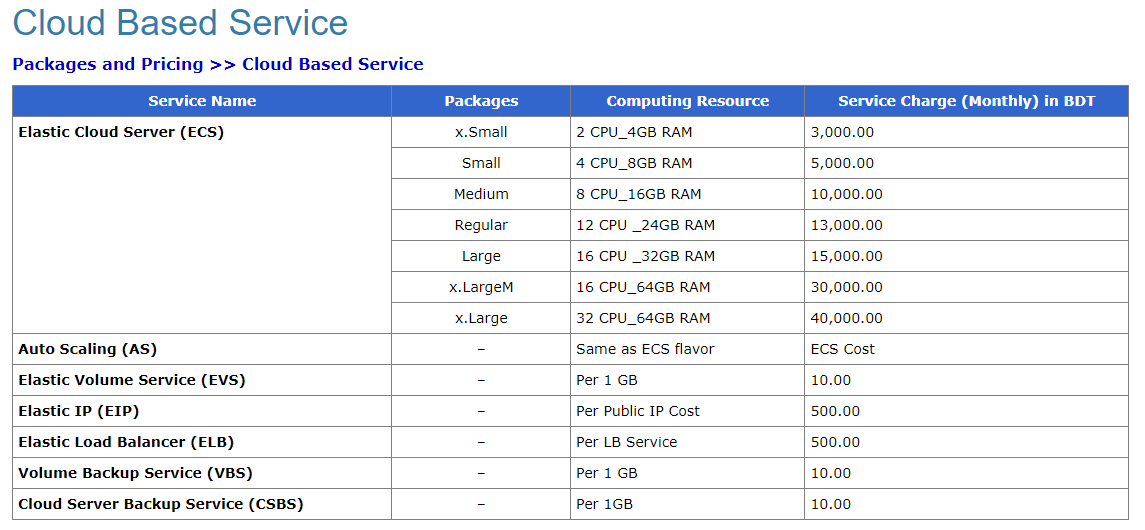}
    \caption{Cloud Service Packages}
    \label{fig:cloudServices}
\end{figure}

As previously stated, we cannot use a third-party cloud service for hosting if the service was developed specifically for government use. The National Data Center (NDC) will provide this service \cite{ndcb_cloud_service}. The Figure \ref{fig:cloudServices} shows the package information which is the best fit for our service. As this service is intended for use on a national scale, we will use the X.Large package. Which service is an Elastic Cloud (ECS). ECS is an excellent private blockchain package. With the aid of ECS, we can add any number of new organizations to our network at any time and scale them in 5 or 6 minutes from the server-side. And after purchasing a single Public IP, we can subnet mask these IPs and scale any number of organization peers in the ECS environment using private IPs. In this package, the Load Balancing service is also very inexpensive.

The entire cost estimate was based on the Bangladeshi Taka in May 2022, when 1 USD was worth 86.07 BDT.

\section{Conclusion}
Health care in Bangladesh is a sector in which both governmental and nongovernmental organizations collaborate. Although the quality of services in the private sector is improving day by day and a competitive environment has been created to deliver good services, public hospitals are slowing down for a number of reasons. In many cases, corruption is responsible. Although the government takes various initiatives every year to provide various health services to the people of the country, there is no way to ensure that they reach the people. Because our health care system operates on a national level, the ministry still monitors them in a traditional way. As a result, many wrongdoers get through here easily and without any irregularities. On the other hand, the complaints of the harmed people do not reach the policy makers. Therefore, this study develops a healthcare application framework based on distributed ledger technology that can eliminate corruption and establish a proper healthcare  providing and monitoring system for the citizens of Bangladesh. The proposed system is a simple and reliable digital process for all organizations where all agreements and information are guaranteed to be immutable.

% \section*{Acknowledgement}
% Lorem ipsum dolor sit amet, consectetur adipiscing elit, sed do eiusmod tempor incididunt ut labore et dolore magna aliqua. Ut enim ad minim veniam, quis nostrud exercitation ullamco laboris nisi ut aliquip ex ea commodo consequat. Duis aute irure dolor in reprehenderit in voluptate velit esse cillum dolore eu fugiat nulla pariatur. Excepteur sint occaecat cupidatat non proident, sunt in culpa qui officia deserunt mollit anim id est laborum.

\bibliographystyle{IEEEtran}
\bibliography{access}

\clearpage

\begin{IEEEbiography}[{\includegraphics[width=1in,height=1.25in,clip,keepaspectratio]{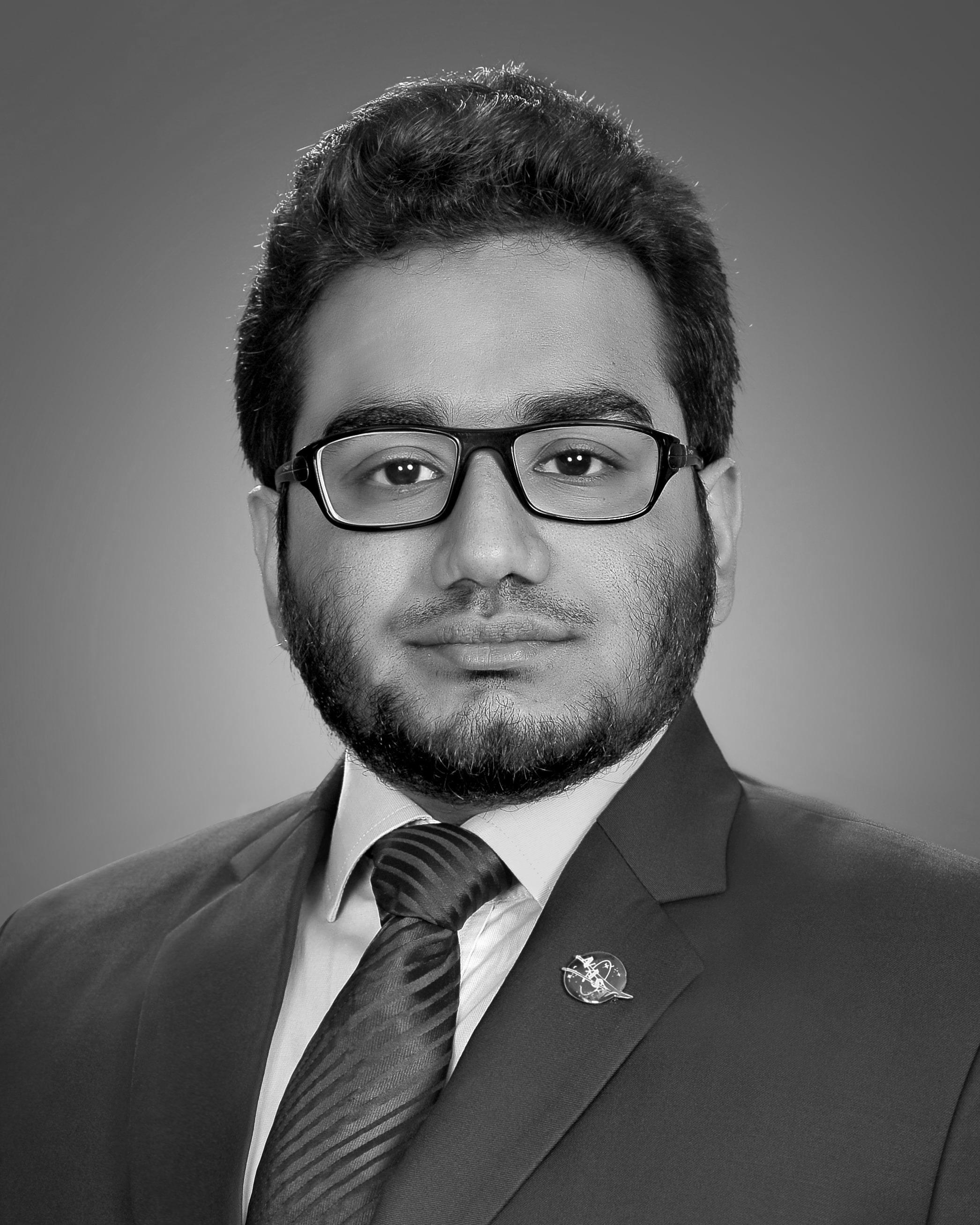}}]{Md. Ariful Islam} received his Bachelor degree in Software Engineering from American International University-Bangladesh (AIUB) in 2021. After three months of internship at EXIM Bank's IT Division he started his professional career with "Robust Research And Development" (RRAD) for Bangladesh Customs and Vat mobile application project (Contract). He also volunteered at one of the international client meeting of "Datasoft Manufacturing and Assembly Inc Limited" as a consultant of Private-Blockchain Technology. In February 2021 he was among the top 40 National Finalist at "Blockchain Olympiad Bangladesh" (BCOLBD) and in its order he was among the top 12 (twelve) of the National Teams of Bangladesh at "International Blockchain Olympiad 2021" (IBCOL 2021). There he have  secured the "Merit Award" on behalf of Bangladesh. After publishing his first work at google play store he got the opportunity from RRAD to design and develop an entire Import Entitlement System for "Bangladesh Customs Bond Commissionerate". Currently he is continuing his career at "Brain Station 23 Ltd" as an Associate Automation Engineer.
\end{IEEEbiography}

\begin{IEEEbiography}[{\includegraphics[width=1in,height=1.25in,clip,keepaspectratio]{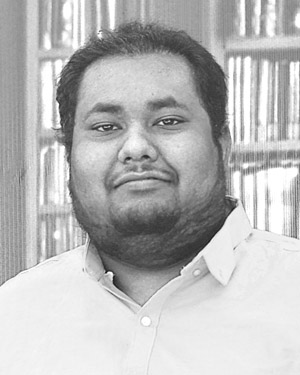}}]{MD. ANTONIN ISLAM} received his Bachelor degree in Software Engineering from American International University-Bangladesh (AIUB) in 2021. After several contract work with domestic and foreign clients he started his professional career with "Robust Research And Development" (RRAD). Before that he also volunteered
at one of the international client meeting of "Data-
soft Manufacturing and Assembly Inc Limited" as
a consultant of Private-Blockchain Technology. In February 2021 his team was among the Top 40 National Finalist at "Blockchain Olympiad Bangladesh" (BCOLBD). He also received the "Merit Award" at "International Blockchain Olympiad 2021" (IBCOL 2021) where his team represented his country as one of Top 12 National Teams of Bangladesh. His first paper "An automated monitoring and environmental control system for laboratory-scale cultivation of oyster mushrooms using the Internet of Agricultural Thing (IoAT)" was presented at the International Conference on Computing Advancement (ICCA 2022). Also, his research interests are in the area of Blockchain, Internet of Things and Cloud Computing.    
\end{IEEEbiography}
\begin{IEEEbiography}[{\includegraphics[width=1in,height=1.25in,clip,keepaspectratio]{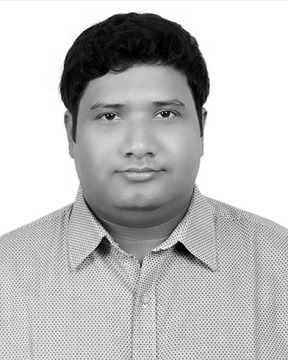}}]{MD. AMZAD HOSSAIN JACKY } received his Bachelor’s degree in Software Engineering (SE) from American International University-Bangladesh (AIUB), Dhaka, Bangladesh in 2021. He is currently working as a software engineer in Star IT Ltd. He was among the top 40 National Finalist at "Blockchain Olympiad Bangladesh"(BCOLBD) 2021. After that he was merit awarded in the "International Blockchain Olympiad 2021". There he worked extensively on server and network configuration and administration. He has done research in the field of Human Computer Interaction (HCI) named “Ball Game Controller: A Tangible User Interface”. Also, His research interests lie in the area of Blockchain Technology, Web 3.0, and Cloud Computing.
\end{IEEEbiography}
\begin{IEEEbiography}[{\includegraphics[width=1in,height=1.25in,clip,keepaspectratio]{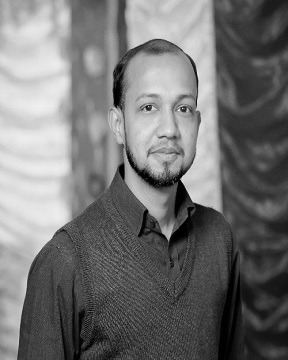}}]{Md. Al-Amin} currently working as a lecturer in the Computer Science Department, American International University-Bangladesh (AIUB). Besides his teaching profession, he is actively doing R\&D projects on freelance platforms for different clients across the globe. He is also supervising research \& development teams \& providing technical consultancy.
He received his Bachelor's in Software Engineering and Master's degree in Computer Science and Engineering (CSE) from American International University-Bangladesh(AIUB), Dhaka, Bangladesh in 2015 and 2017 respectively. During his Master's degree, he was awarded the academic distinction Magna Cum Laude(Silver Medal) award for his academic results. He achieved two times ICT Fellowship Awards for years 2015-16 \& 2016-17 from the Ministry of ICT, Government of Bangladesh.
He has also served as a registration committee member in the International Conference on Computing Advancements (ICCA 2020). He is a member of Bangladesh's computer society.
His research interest primarily focuses on Distributed Ledger Technology (DLT), Blockchain Technology, Web 3.0, Distributed Computing, Web of Things, Information Security, Web Assembly, and Knowledge base System.
\end{IEEEbiography}
\begin{IEEEbiography}[{\includegraphics[width=1in,height=1.25in,clip,keepaspectratio]{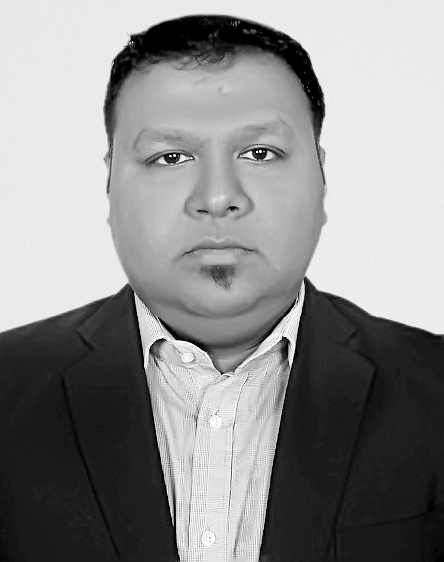}}]{M. Saef Ullah Miah} is a PhD candidate at University Malaysia Pahang (UMP) and is currently working as a Graduate Research Assistant at the Faculty of Computing, UMP. He was an assistant professor in the Department of Computer Science, American International University-Bangladesh (AIUB). He is currently engaged in research and teaching activities and has practical experience in software development and project management. He earned his Master of Science and Bachelor of Science degrees from AIUB. In addition to his professional activities, he is passionate about working on various open source projects. His main research interests are data and text mining, natural language processing, machine learning, material informatics and blockchain applications.
\end{IEEEbiography}
\begin{IEEEbiography}[{\includegraphics[width=1in,height=1.25in,clip,keepaspectratio]{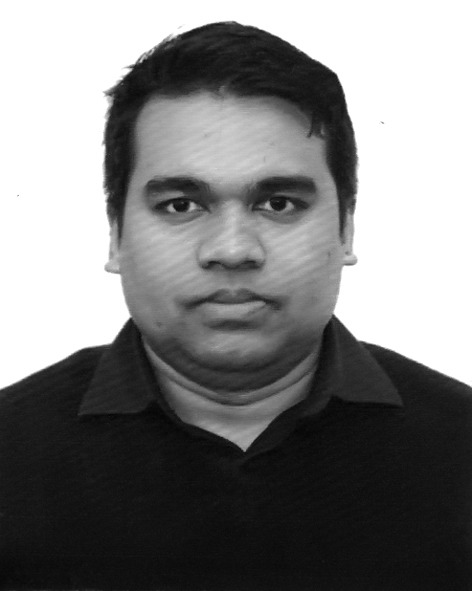}}]{Md Muhidul Islam Khan} received his Bachelor degree in Computer Science and Engineering from Khulna University of Engineering and Technology (KUET) in 2007. In 2009 he received his Masters degree from Bangladesh University of Engineering and Technology (BUET). He has participated in the "eLINK"-project at Corvinus University of Budapest, Hungary, from September 2009 until July 2010 (funded by the European Union). His specialization lies in the fields of Wireless Sensor Networks, Networked Embedded Systems, Pervasive Computing, and Machine learning/AI-based resource allocation in Wireless Networks, blockchain-based technologies. He completed his PhD in 2014 under the Erasmus Mundus Grant from the European Commission working at Klagenfurt University, Austria, and University of Genova, Italy. He joined as an Assistant Professor at BRAC University, Bangladesh, and served there for one year. After that, he completed his one-year postdoc from the Hebei University of Technology, Tianjin, China. He worked as a Research Scientist in the Electronics Department at Tallinn University of Technology, Estonia. He worked as a senior lecturer in the School of Information Technologies, Tallinn University of Technology, Tallinn, Estonia. Currently, he is working as an Associate Researcher at the University of Stavanger, Norway.
\end{IEEEbiography}

\begin{IEEEbiography}[{\includegraphics[width=1in,height=1.25in,clip,keepaspectratio]{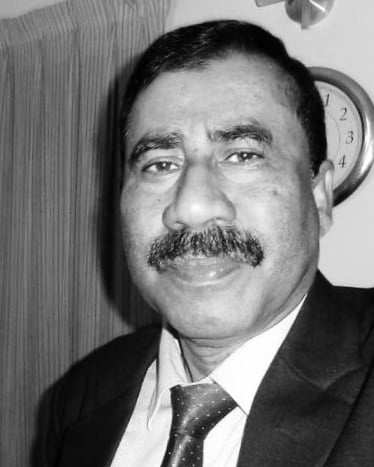}}]{Md. Iqbal Hossain}, a Freedom Fighter of the Bangladesh Liberation War (1971) \& after the war a Doctor who earned his MBBS degree from Chittagong Medical College under Chittagong University on 1981. He start his career at Rural Health complex from 1983 to 1989 as a medical officer. From that time he was working for the health of poor Rural people of Bangladesh. Also took part in the EPI program. On 1994 he received his Post-graduation degree on Pediatric from IPGMR, Dhaka and Bangladesh College of Physician and Surgeon. From 1995 he start his work as  consultant Pediatrics at Rangamati General Hospital till 1999. On 1999 he join at Mymensing Medical College as Asst. Professor on Pediatric. And worked there till 2000.  After that he went Saudi Arabia and worked in Al-Jouf Maternity and Children Hospital  till 2005. On 2005 he return Back to Bangladesh and join at Faridpur Medical College as  Asst. Professor.  During this time he gave consultancy Service and teaching on  pediatrics to the Undergraduate Medical Student. He worked there till 2010 and then he left Bangladesh again and served Maldives health service till 2015 as pediatrician. He is currently serving the people of his own country in this pandemic situation (COVID-19) at a private hospital in Chakaria, Cox's Bazar district, Bangladesh.
\end{IEEEbiography}

\EOD

\end{document}